\documentstyle[12pt,graphics,epsfig,wrapfig]{article}
\newcommand{\be}{\begin{equation}}
\newcommand{\ee}{\end{equation}} 
\newcommand{\bea}{\begin{eqnarray}}
\newcommand{\eea}{\end{eqnarray}}
\newcommand{\la}{\langle}
\newcommand{\ra}{\rangle}
\newcommand{\di}{ {\rm d} }

\newcommand{\SIDIS}{{\mbox{\tiny SIDIS}}}
\newcommand{\DY   }{{\mbox{\tiny DY}}}
\newcommand{\aH}{ {a} }
\begin{document}
\begin{center}
{\bfseries SIVERS AND COLLINS SINGLE SPIN ASYMMETRIES}

\vskip 5mm
\underline{A.~V.~Efremov}$^{1\dag}$, {K.~Goeke}$^{2}$ and
P.~Schweitzer$^{2}$

\vskip 5mm
{\small
(1) {\it
Joint Institute for Nuclear Research, Dubna, 141980 Russia
}\\
(2) {\it
Institut f\"ur Theoretische Physik II, Ruhr-Universit\"at
Bochum, Germany
}\\
$\dag$ {\it
E-mail: efremov@theor.jinr.ru
}}
\end{center}

\begin{abstract}
The Sivers and Collins asymmetries are 
the most prominent Single Spin Asymmetries (SSA)
in Semi-Inclusive 
Deeply Inelastic Scattering (SIDIS)
with transverse target polarization. 
In this talk we present our understanding of these phenomena.
\end{abstract}

\section{Introduction}

SSAs in hard reactions have a long 
history dating back to the 1970s when significant polarizations of 
$\Lambda$-hyperons in collisions of unpolarized hadrons were 
observed \cite{Bunce:1976yb}, and to the early 1990s when large 
asymmetries in $p^\uparrow p\to\pi X$ or $p^\uparrow \bar{p}\to\pi 
X$ were found at Protvino \cite{Apokin:1988sn} and FNAL 
\cite{Adams:1991rw}. No fully consistent and 
satisfactory unifying approach to the theoretical description of 
these observations has been found so far --- see the review 
\cite{Anselmino:2002mx}.

Interestingly, the most recently observed SSA and azimuthal phenomena, 
namely those in SIDIS and $e^+e^-$ annihilations seem
better under control. This is in particular the case for the
transverse target SSA observed at HERMES and COMPASS
\cite{Airapetian:2004tw,Alexakhin:2005iw,Diefenthaler:2005gx}
and the azimuthal correlations in hadron production in 
 $e^+e^-$ annihilations observed at BELLE \cite{Abe:2005zx}. 
On the basis of a generalized factorization approach in which
transverse parton momenta are taken into account
\cite{Ji:2004wu} these ``leading
twist'' asymmetries can be explained \cite{Mulders:1995dh,Boer:1997mf} 
in terms of the Sivers
\cite{Sivers:1989cc,Brodsky:2002cx,Collins:2002kn,Belitsky:2002sm}
or Collins effect \cite{Collins:1992kk}. The former describes,
loosely speaking, the distribution of unpolarized partons in a
transversely polarized proton, the latter describes a left-right
asymmetry in
fragmentation of transversely polarized partons into unpolarized
hadrons. In the transverse target SSA these effects can be
distinguished by the different azimuthal angle distribution of the
produced hadrons: Sivers  effect $\propto\sin(\phi-\phi_S)$, while
Collins effect $\propto\sin(\phi+\phi_S)$, where $\phi$ and
$\phi_S$ denote respectively the azimuthal angles of the produced
hadron and the target polarization vector with respect to the axis
defined by the hard virtual photon \cite{Mulders:1995dh}. 
Both effects have been subject to intensive phenomenological studies in
hadron-hadron-collisions
\cite{Anselmino:2004ky} and in SIDIS
\cite{Efremov:2004tp}-\cite{Efremov:2007kj}.
In this talk our understanding of these phenomena is presented.

For the longitudinal target SSA in SIDIS, which were observed
first \cite{Airapetian:1999tv,Avakian:2003pk}
but are dominated by subleading-twist effects
\cite{Efremov:2001cz,Efremov:2001ia}, the situation
is less clear and their description ({\sl presuming} 
factorization holds) is more involved.

\section{Sivers effect}
The Sivers effect \cite{Sivers:1989cc} was originally suggested
to explain the large SSAs in 
$p^\uparrow p\to\pi X$ (and $\bar{p}^\uparrow p\to\pi X$) observed 
at FNAL \cite{Adams:1991rw} and confirmed at higher energies 
by RHIC \cite{Adams:2003fx}. 
It is due a correlation between (the transverse component of) the
nucleon spin ${\bf S}_{\rm T}$ and intrinsic transverse parton
momenta ${\bf p}_{\rm T}$ in the nucleon, and decribed by the
Sivers function $f_{1T}^\perp(x,{\bf p}_T^2)$
whose precise definition in QCD was worked out only recently
\cite{Collins:2002kn,Belitsky:2002sm}.

\begin{wrapfigure}[9]{R}{.41\textwidth}
\begin{center}
\vskip-8mm
\includegraphics[width=.38\textwidth]{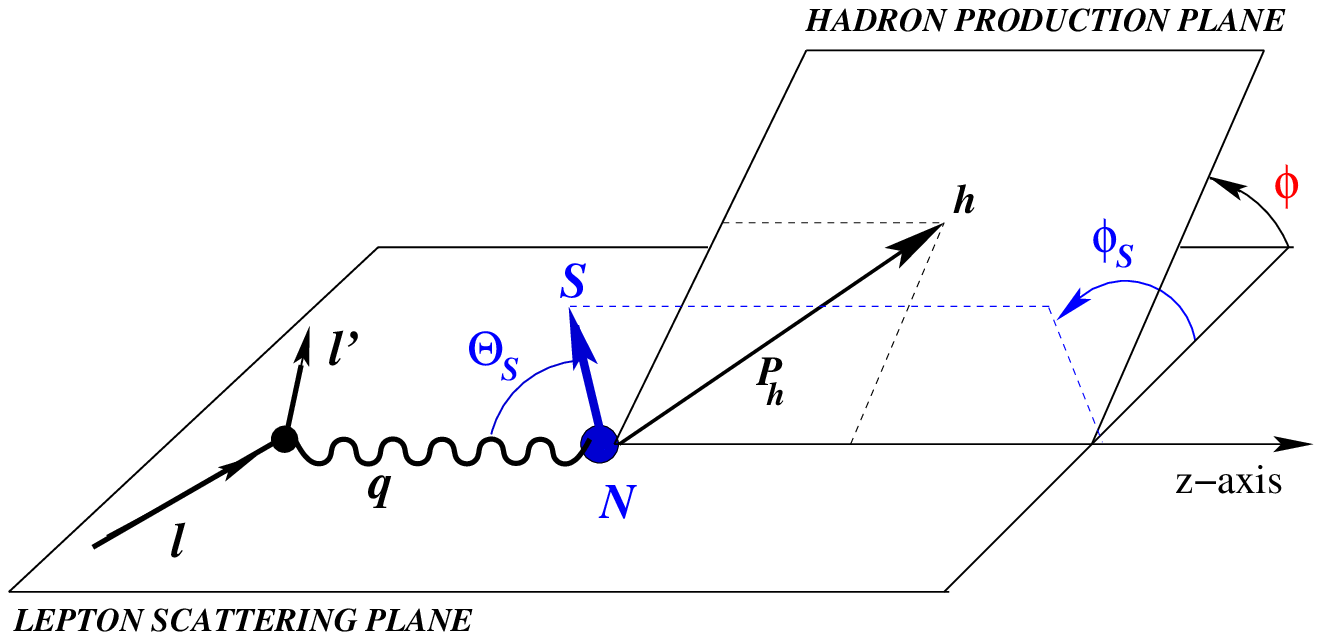}
\end{center}
\vskip-5mm
\caption{\label{fig2-processes-kinematics}\footnotesize 
Kinematics of the SIDIS process $lN\rightarrow l^\prime h X$. }
\end{wrapfigure}

\paragraph{2.1 Sivers effect in SIDIS.}
The azimuthal SSA measured by HERMES \& COMPASS in the SIDIS 
process $lp^\uparrow\rightarrow l'h X$ (see 
Fig.~\ref{fig2-processes-kinematics}) is defined as
\be
\frac{N^\uparrow -N^\downarrow}{N^\uparrow+ N^\downarrow}\propto
\underbrace{\sin(\phi-\phi_S)\;A_{UT}^{\sin(\phi-\phi_S)}}_{\hspace{1cm} \mbox{Sivers
\hspace{0.3cm} and}\hspace{-1.5cm}} +
\underbrace{\sin(\phi+\phi_S)\;A_{UT}^{\sin(\phi+\phi_S)}}_{\hspace{1cm}
\mbox{Collins effect}}
\ee

\noindent 
where $N^{\uparrow(\downarrow)}$ are the event counts 
for the respective transverse target  polarization. We assume the 
distributions of transverse parton and hadron momenta in 
distribution (DF) and fragmentation function (FF) to be Gaussian 
with corresponding averaged transverse momenta,
$p^2_{\rm Siv}$ and $K^2_{\!D_1}$, taken $x$- or $z$-
and flavour-independent. The Sivers SSA as measured in 
\cite{Airapetian:2004tw,Alexakhin:2005iw} is then given 
by \cite{Collins:2005ie}
\be
\label{Eq:AUT-SIDIS-Gauss}
\hskip-3mm
        A_{UT}^{\sin(\phi-\phi_S)} = (-2)\; \frac{a_{\rm G}
        \sum_a e_a^2\,x f_{1T}^{\perp(1)a}(x)\, D_1^{a}(z)}{
        \sum_a e_a^2\,x f_1^a(x)\,D_1^{a}(z)}
    \;\;\;\mbox{with}\;\,\; 
    a_{\rm G}=\frac{\sqrt{\pi}}{2}
    \frac{M_N}{\sqrt{p^2_{\rm Siv}+K^2_{\!D_1}/z^2}}
\ee
and 
$f_{1T}^{\perp(1)a}(x)
\equiv\int\!\di^2{\bf p}_T \frac{{\bf p}_T^2}{2 M_N^2}
f_{1T}^{\perp a}(x,{\bf p}_T^2)$.
In the limit a large number of colours $N_c$ one has
\be\label{Eq:large-Nc}
      f_{1T}^{\perp u}(x,{\bf p}_T^2) =
    - f_{1T}^{\perp d}(x,{\bf p}_T^2) \;\;\;
    \mbox{modulo $1/N_c$ corrections,}\ee
and analog for antiquarks
for $x$ of the order $xN_c={\cal O}(N_c^0)$ 
\cite{Pobylitsa:2003ty}. In the following effects of 
antiquarks and heavier flavours are neglected.
%
%
\begin{figure}[b!]
\vspace{-5mm}
\begin{tabular}{ccc}
\includegraphics[width=0.28\textwidth]{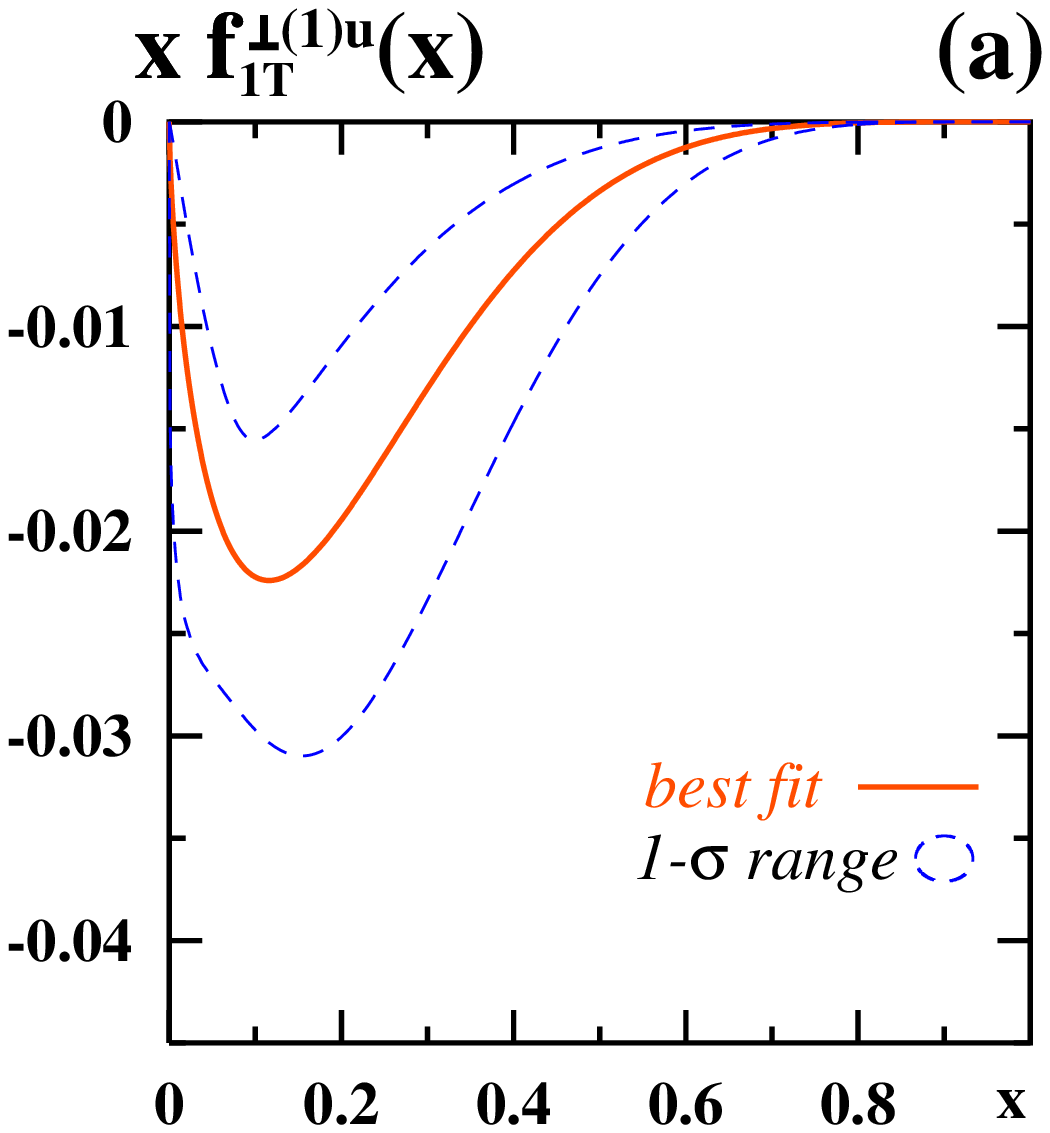}&
\includegraphics[width=0.28\textwidth]{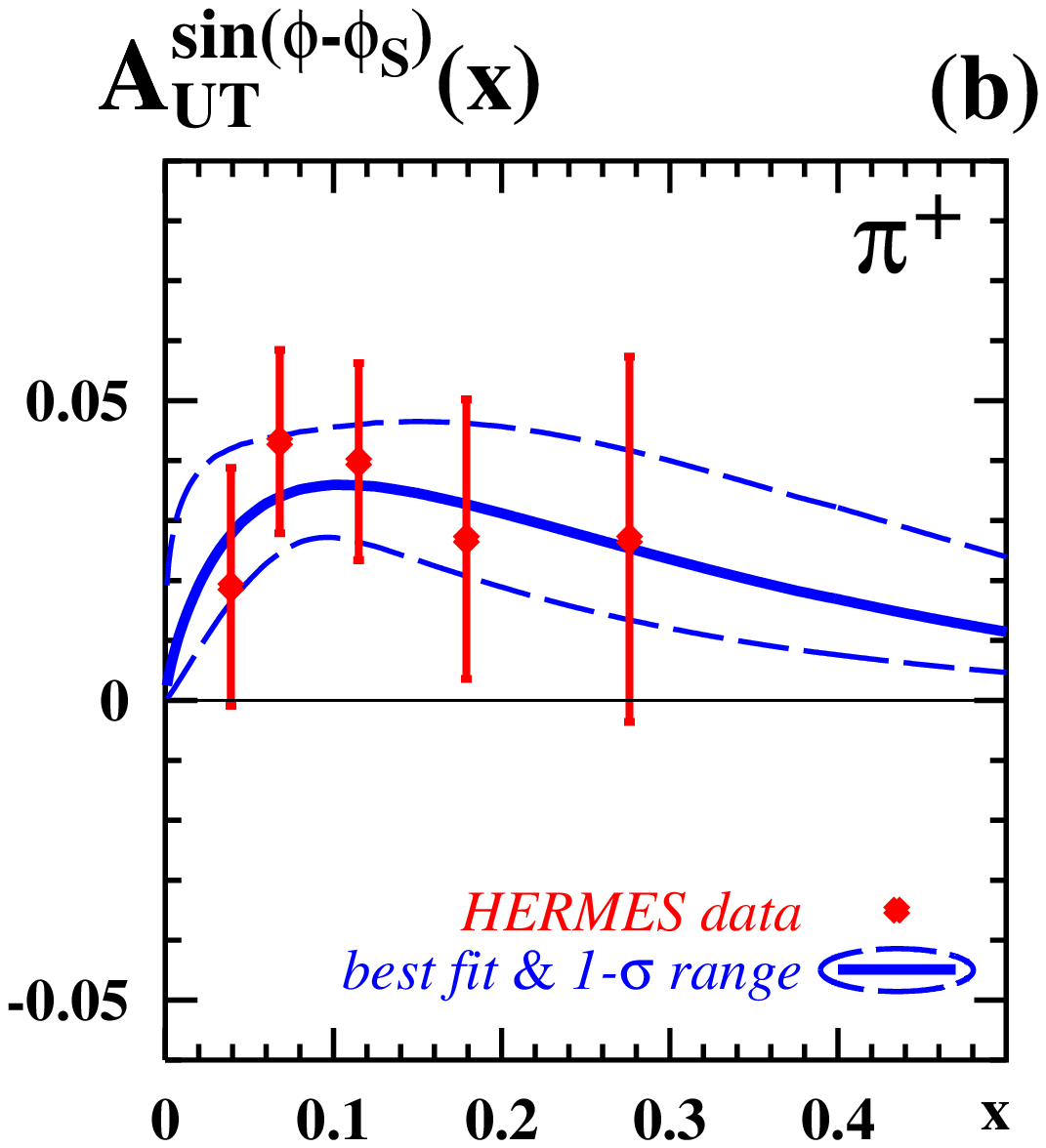}&
\includegraphics[width=0.28\textwidth]{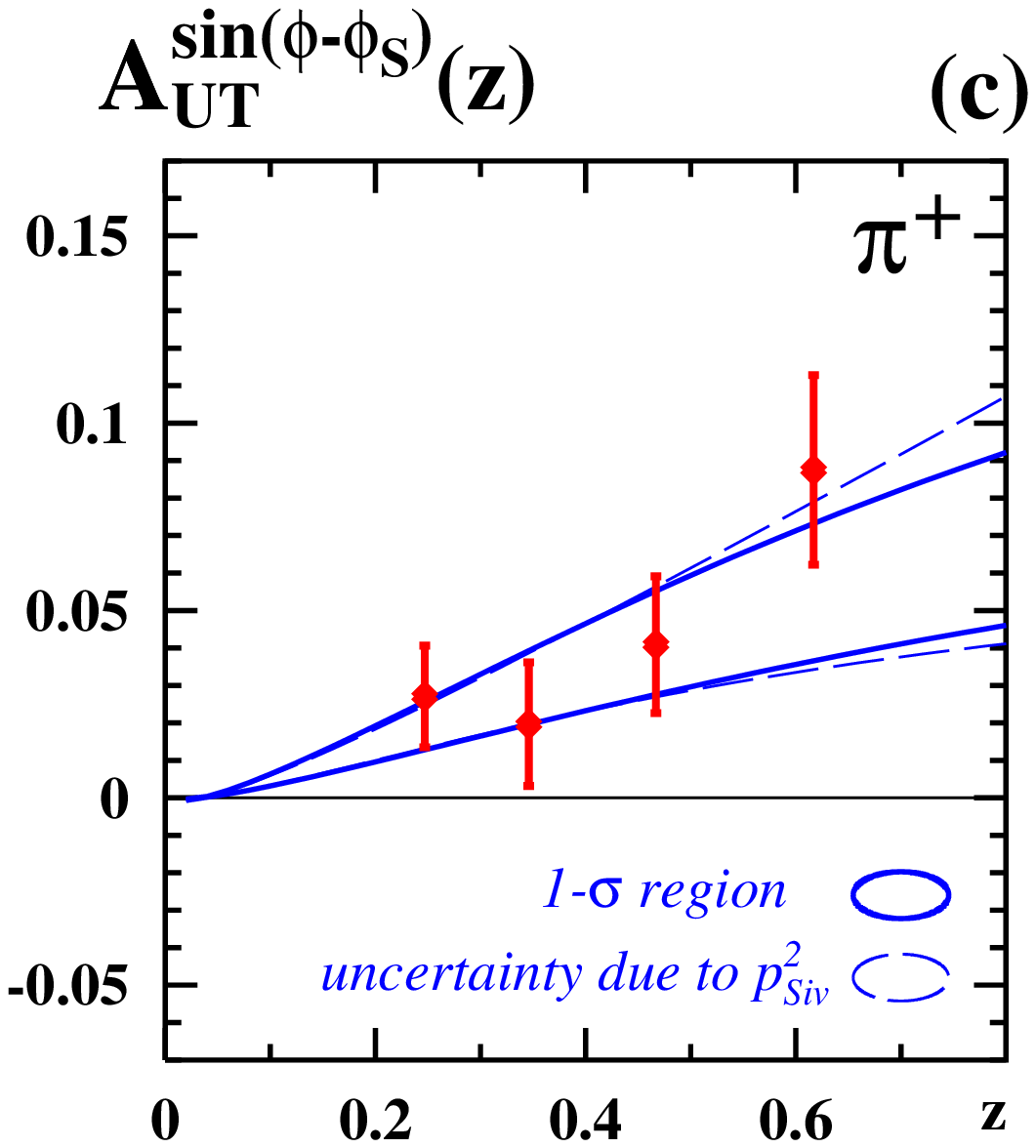}
\end{tabular}
\vspace{-4mm}
\caption{\label{Fig5-compare-to-data}\footnotesize
        {\bf a.} The $u$-quark Sivers function vs.\  $x$ at 
    a scale of $2.5\,{\rm GeV}^2$, as obtained from the
        HERMES data \cite{Airapetian:2004tw}. Shown are the best
        fit and its 1-$\sigma$ uncertainty.
        {\bf b.} and {\bf c.} The azimuthal SSA
        $A_{UT}^{\sin(\phi_h-\phi_S)}$ as function of $x$ and $z$
        for positive pions as obtained from the fit
        (\ref{Eq:ansatz+fit}) in comparison to the
        data \cite{Airapetian:2004tw}.}
\end{figure}
%
%
It was shown \cite{Collins:2005ie} that 
the large-$N_c$ relation (\ref{Eq:large-Nc})
describes the HERMES data \cite{Airapetian:2004tw} 
by the following 2-parameter Ansatz and best fit
\be\label{Eq:ansatz+fit}
        xf_{1T\SIDIS}^{\perp (1) u}(x) = -xf_{1T\SIDIS}^{\perp (1) d}(x)
        \stackrel{\rm Ansatz}{=}\;  A \, x^b   \,(1-x)^5
        \,\,\stackrel{\rm fit}{=}\,\, -0.17 x^{0.66}(1-x)^5\;.
\ee
Fig.~\ref{Fig5-compare-to-data}a shows
the fit and its 1-$\sigma$ uncertainty due to the
statistical error of the data \cite{Airapetian:2004tw}.
Fig.~\ref{Fig5-compare-to-data}b shows that this fit very well 
describes the $x$-dependence of the HERMES data \cite{Airapetian:2004tw}.
Fig.~\ref{Fig5-compare-to-data}c finally shows the equally good
description of the $z$-dependence of the data \cite{Airapetian:2004tw}
that were not included in the fit, and serves here as a cross check 
for the Gauss Ansatz.

We have explicitly checked that effects due to Sivers $\bar u$- and 
$\bar d$-distributions cannot be resolved within the error bars of 
the data \cite{Airapetian:2004tw} (however, see Sec.~\ref{Sec:new-results}).
We also checked that $1/N_c$-corrections are within
the error bars of the data \cite{Airapetian:2004tw}.
For that we assumed that the
flavour singlet Sivers distribution is 
suppressed by exactly a factor of $1/N_c$ with respect to the
flavour non-singlet combination according to
Eq.~(\ref{Eq:large-Nc}). That is, with $N_c=3$,
\be\label{Eq:model-Nc-corr}
\Big |(f_{1T}^{\perp(1)u}+f_{1T}^{\perp(1)d})(x) \Big | 
\stackrel{!}{=} \pm \;\frac{1}{N_c} 
(f_{1T}^{\perp(1)u}-f_{1T}^{\perp(1)d})(x) \,,
\ee

%
\begin{wrapfigure}[14]{HR}{.35\textwidth}
\begin{center}
\vskip-7mm
\includegraphics[width=.35\textwidth]{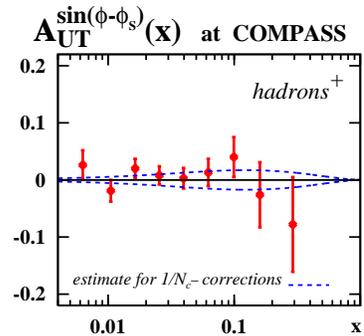}
\vskip-4mm
\begin{minipage}{.3\textwidth}
\caption{\label{Fig9-AUT-Nc-COMPASS}\footnotesize
        The Sivers SSA for positive hadrons from deuteron.
    Data are from COMPASS \cite{Alexakhin:2005iw}.
    The theoretical curves indicate the magnitude of the effect 
    on the basis of the estimate (\ref{Eq:model-Nc-corr}).}
\end{minipage}
\end{center}
\end{wrapfigure}
\noindent 
where we use $f_{1T}^{\perp(1)q}(x)$ from (\ref{Eq:ansatz+fit})
on the right-hand-side.

On an isoscalar target, such as deuteron, the entire effect 
is due to $1/N_c$-corrections.
Assuming that charged hadrons at COMPASS are mainly 
pions, the rough estimate (\ref{Eq:model-Nc-corr}) of $1/N_c$-corrections 
yields results compatible with the COMPASS data \cite{Alexakhin:2005iw},
see Fig.~\ref{Fig9-AUT-Nc-COMPASS}. 

Thus, the large-$N_c$ approach works, because the precision of the 
first data \cite{Airapetian:2004tw,Alexakhin:2005iw} is comparable 
to the theoretical accuracy of the large-$N_c$ relation
(\ref{Eq:large-Nc}). 
Our results are in agreement with other studies
\cite{Anselmino:2005ea,Vogelsang:2005cs,Anselmino:2005an}.

We conclude that the HERMES and COMPASS data
\cite{Airapetian:2004tw,Alexakhin:2005iw} are compatible 
with the large-$N_c$ prediction (\ref{Eq:large-Nc}) for the 
Sivers function \cite{Pobylitsa:2003ty}.
Remarkably, the sign of the extracted Sivers function in
Eq.~(\ref{Eq:ansatz+fit}) agrees with the physical picture
discussed in \cite{Burkardt:2002ks}.

\paragraph{2.2 Sivers effect in the Drell-Yan process.}
Universility is a particularly interesting aspect of the Sivers 
function. On the basis of time-reversal arguments it is predicted
\cite{Collins:2002kn} that this (and other ``T-odd'') distribution(s)
have opposite signs in SIDIS and DY
\be\label{Eq:01}
     f_{1T}^\perp(x,{\bf p}_T^2)_\SIDIS =
    -f_{1T}^\perp(x,{\bf p}_T^2)_\DY \;. \ee The experimental check
of Eq.~(\ref{Eq:01})  would provide a thorough test of our understanding of
the Sivers effect within QCD. In particular, the experimental
verification of (\ref{Eq:01}) is a crucial prerequisite for testing
the factorization approach to the description of processes containing
$p_T$-dependent correlators \cite{Ji:2004wu}.

On the basis of the first information of the Sivers effect in SIDIS
\cite{Airapetian:2004tw,Alexakhin:2005iw} it was shown that the 
Sivers effect leads to sizeable SSA in $p^\uparrow\pi^-\to 
l^+l^-X$, which could be studied at COMPASS, and in 
$p^\uparrow\bar{p}\to l^+l^-X$ or $p\bar{p}^\uparrow\to l^+l^-X$ 
in the planned PAX experiment at GSI \cite{PAX}
making the experimental check of Eq.~(\ref{Eq:01}) feasible and 
promising \cite{Efremov:2004tp}. Both experiments are
dominated by annihilations of valence quarks (from $p$) and 
valence antiquarks (from $\bar{p}$, $\pi^-$). This yields 
sizeable counting rates, and the processes are not 
sensitive to Sivers antiquarks, that are not 
constrained by the present data,
see \cite{Efremov:2004tp}-\cite{Collins:2005ie}.

On a shorter term the Sivers effect in DY can be studied in
$p^\uparrow p\to l^+l^-X$ at RHIC. In $pp$-collisions inevitably 
antiquark distributions are involved, and the counting rates are 
smaller. We have shown, however, that the Sivers SSA in DY can 
nevertheless be measured at RHIC with an accuracy sufficient to 
unambiguously test Eq.~(\ref{Eq:01}) \cite{Collins:2005rq}.

The theoretical understanding of SSA in $p^\uparrow p\to\pi X$, 
which originally motivated the introduction of the Sivers effect, 
is more involved compared to SIDIS or DY. No factorization proof
is formulated for this process. The SSA can also be generated 
by twist-3 effects \cite{Efremov:eb}
that, however, could be manifestations of the same effect 
in different $k_T$ regions \cite{Bomhof:2004aw}.


\section{Transversity and Collins effect}

The transversity distribution function $h_1^a(x)$ enters the 
expression for the Collins SSA in SIDIS 
together with the equally unknown Collins fragmentation function
\cite{Collins:1992kk} (FF) $H_1^a(z)$\footnote{
    \label{Footnote-1}
    We assume a factorized Gaussian dependence on parton and
    hadron transverse momenta \cite{Mulders:1995dh} with
    $B_{\rm G}(z)=(1+z^2\;\la{\bf p}_{h_1}^2\ra/\la{\bf K}^2_{H_1}\ra)^{-1/2}$
    and define $H_1^\aH(z) \equiv H_1^{\perp (1/2) a}(z)= $
    $\int\di^2{\bf K}_T\frac{|{\bf K}_T|}{2zm_\pi} H_1^{\perp a}(z,{\bf K}_T)$.
    The Gaussian widths are assumed flavor and $x$- or $z$-independent.
    We neglect throughout the soft factors \cite{Ji:2004wu}.}
\be\label{Eq:AUT-Collins-1}
    A_{UT}^{\sin(\phi+\phi_S)}
    \!= 2\frac{\sum_a e_a^2 x h_1^a(x)B_{\rm G}
    H_1^{a}(z)}{\sum_a e_a^2\,x f_1^a(x)\,D_1^{a}(z)} \;.
\ee

However, $H_1^{a}(z)$ is accessible in $e^+e^-\to \bar q q\to
2{\rm jets}$ where the quark transverse spin correlation induces 
a specific azimuthal correlation of two hadrons in opposite jets 
\cite{Boer:1997mf}
\be\label{Eq:A1-in-e+e}
    \di\sigma=\di\sigma_{\rm unp}\underbrace{\Biggl[
    1+\cos(2\phi_1)\frac{\sin^2\theta}{1+\cos^2\theta}
    \;C_{\rm G}\times
    \frac{\sum_a e_a^2 H_1^{a}H_1^{\bar a}}
    {\sum_a e_a^2 D_1^a D_1^{\bar a}}\Biggr]}_{\equiv A_1}
\ee
where $\phi_1$ is azimuthal angle of hadron 1 around z-axis along
hadron 2, and $\theta$ is electron polar angle. Also here we
assume the Gauss model and $C_{\rm
G}(z_1,z_2)=\frac{16}{\pi}{z_1z_2}/{(z_1^2+z_2^2)}$.

First experimental indications for the Collins effect were
obtained from studies of preliminary SMC data on SIDIS
\cite{Bravar:1999rq} and DELPHI data on charged hadron production
in $e^+e^-$ annihilations at the $Z^0$-pole \cite{Efremov:1998vd}.
More recently HERMES reported data on the Collins (SSA) in SIDIS
from proton target \cite{Airapetian:2004tw,Diefenthaler:2005gx}
giving the first unambiguous evidence that $H_1^a$ and $h_1^a(x)$
are non-zero, while in the COMPASS experiment
\cite{Alexakhin:2005iw} the Collins effect from a deuteron target
was found compatible with zero within error bars. Finally,
year ago the BELLE collaboration presented data on sizeable azimuthal
correlation in $e^+e^-$ annihilations at a center of mass energy
of $60\,{\rm MeV}$ below the $\Upsilon$-resonance \cite{Abe:2005zx}.

The question which arises is: {\sl Are all these data from 
different SIDIS and $e^+e^-$ experiments compatible, i.e. due to 
the same Collins effect?}

In order to answer this question we extract $H_1^a$ from HERMES
\cite{Diefenthaler:2005gx} and BELLE  
\cite{Abe:2005zx} data, and compare the 
ratios $H_1^a/D_1^a$ from these and other experiments. Such
``analyzing powers'' might be expected to be weakly
scale-dependent.

\paragraph{3.1 Collins effect in SIDIS.}
A simultanous extraction of $h_1^a(x)$ and $H_1^{\perp a}(z)$
from SIDIS data is presently not possible.
We use therefore for $h_1^a(x)$ predictions from 
chiral quark-soliton model \cite{Schweitzer:2001sr} which 
provides a good description of $f_1^a(x)$ and $g_1^a(x)$.
The HERMES data on the Collins SSA \cite{Diefenthaler:2005gx}
can be described in this approach if, at $\la Q^2\ra=2.5\,{\rm GeV}^2$,
\be
\label{Eq:Apower-HERMES}
\frac{\la 2 B_{\rm G}H_1^{\rm fav}\ra}{\la D_1^{\rm fav}\ra}
\biggr|_{\rm HERMES}\hspace{-11mm}=(7.2\pm 1.7)\% \;, \;\;\;\; \\
\frac{\la 2 B_{\rm G}H_1^{\rm unf}\ra}{\la D_1^{\rm unf}\ra}
\biggr|_{\rm HERMES}\hspace{-11mm} = -(14.2\pm 2.7)\%\;.
\ee
where ``${\rm fav}$'' (``${\rm unf}$'') means favored $u\to\pi^+$ 
etc.\ (unfavored $u\to\pi^-$, etc.) fragmentation, and 
$\la\dots\ra$ denotes average over $z$ within the HERMES cuts 
$0.2\le z \le 0.7$.

\begin{figure}[ht!]
\begin{center}
\vspace{-12mm}

\hspace{-5mm}
\includegraphics[width=1.5in]{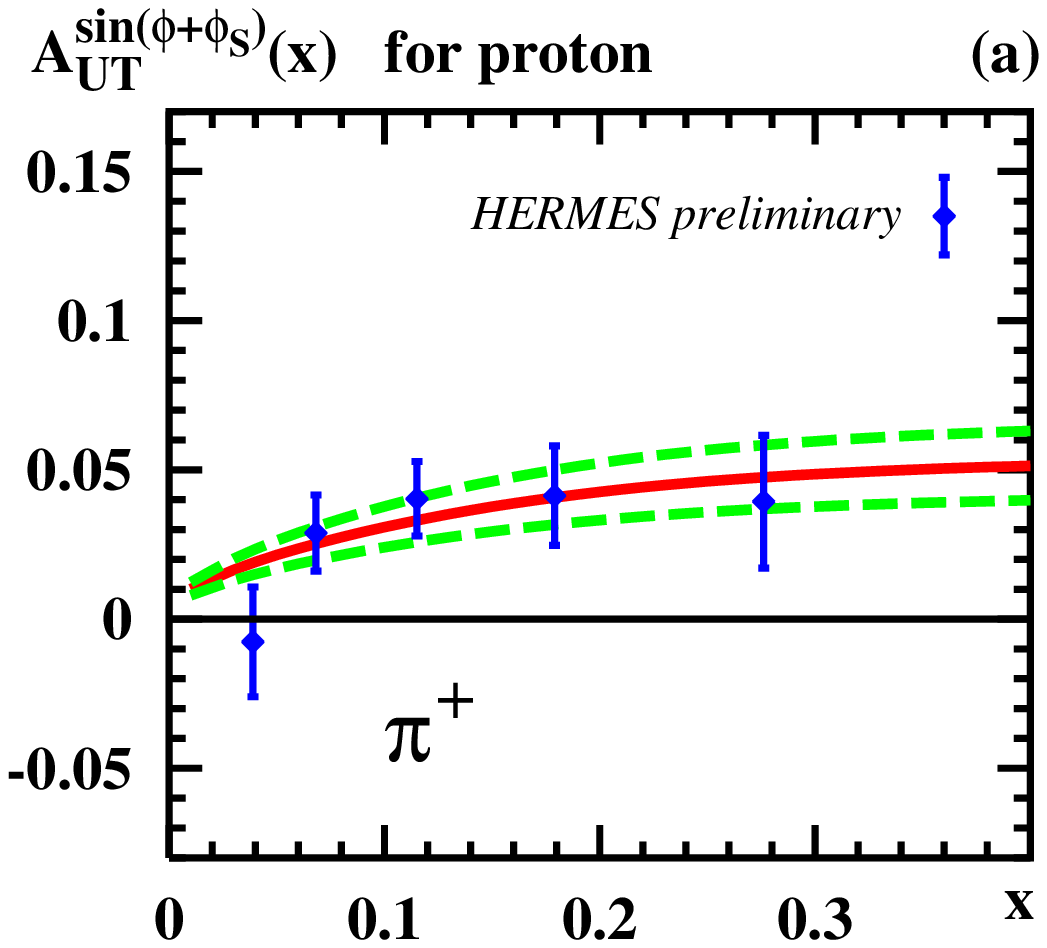}
\hspace{-1mm}
\includegraphics[width=1.5in]{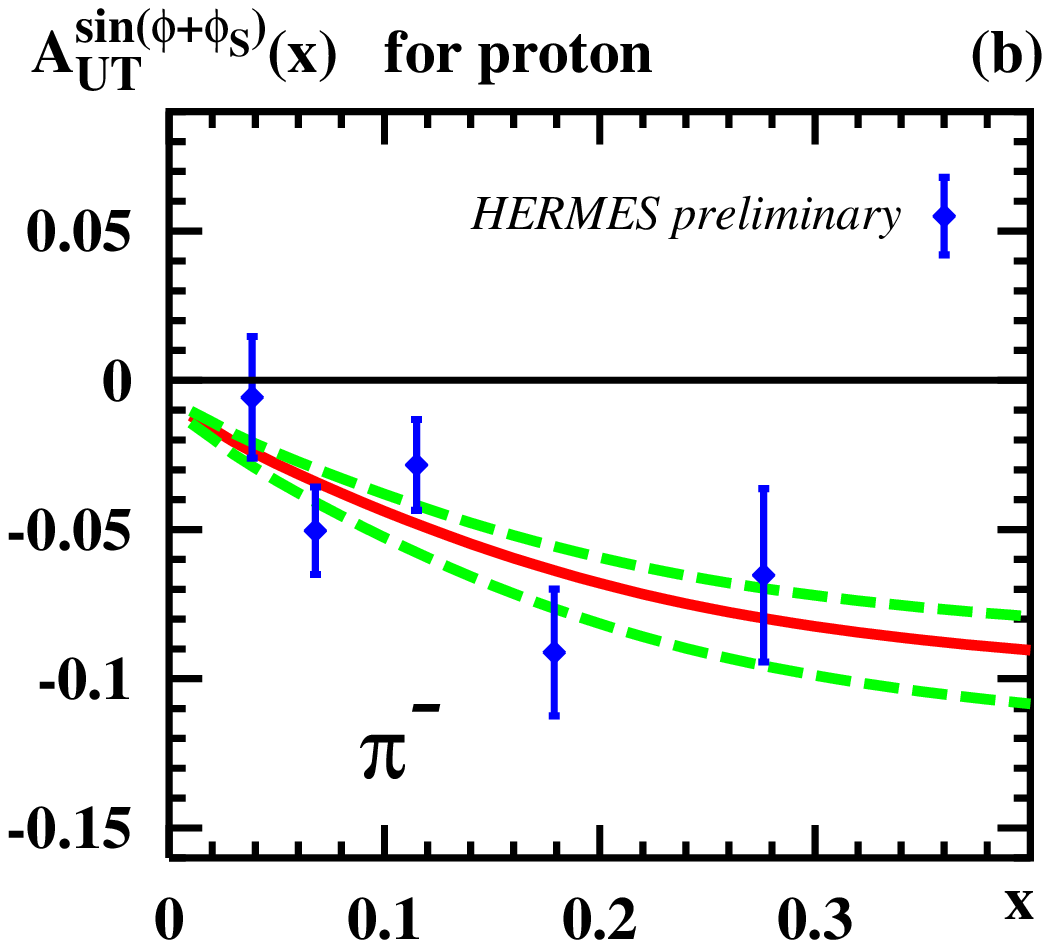}
\hspace{-4mm}
\includegraphics[width=1.6in]{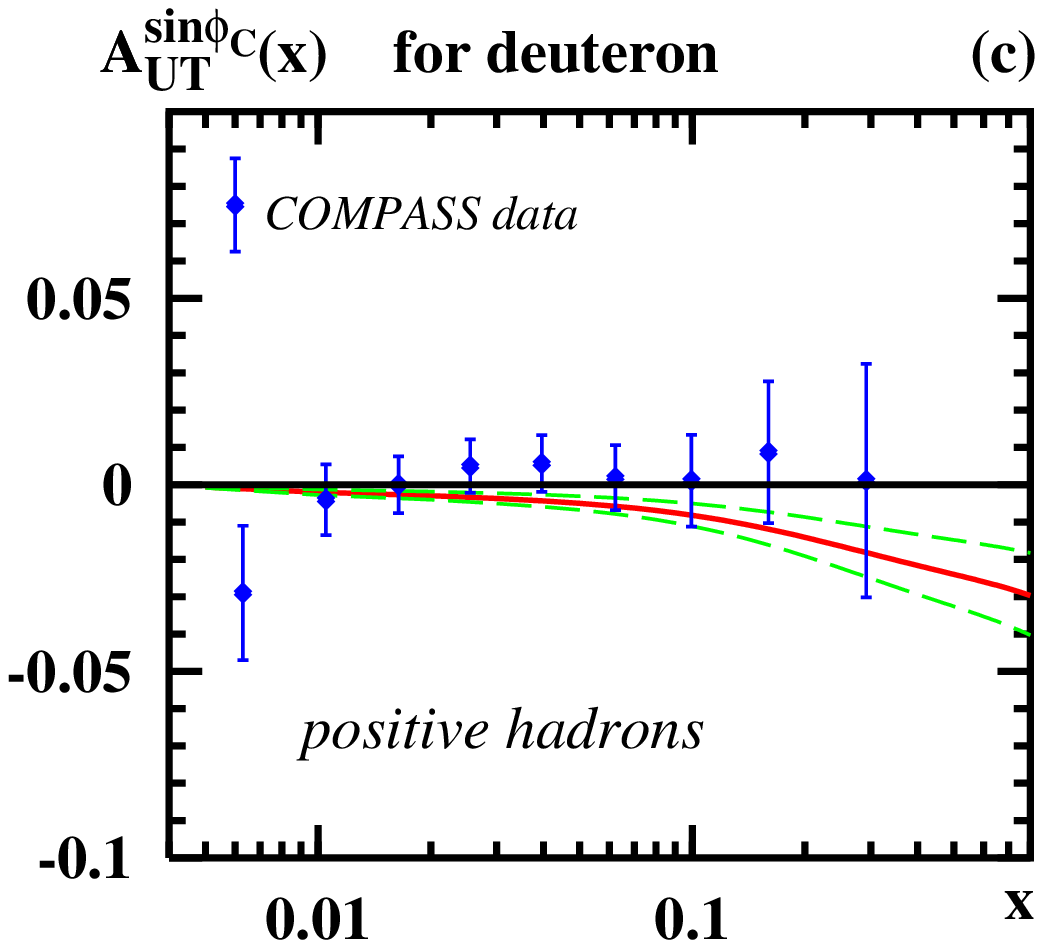} 
\hspace{-4mm}
\includegraphics[width=1.6in]{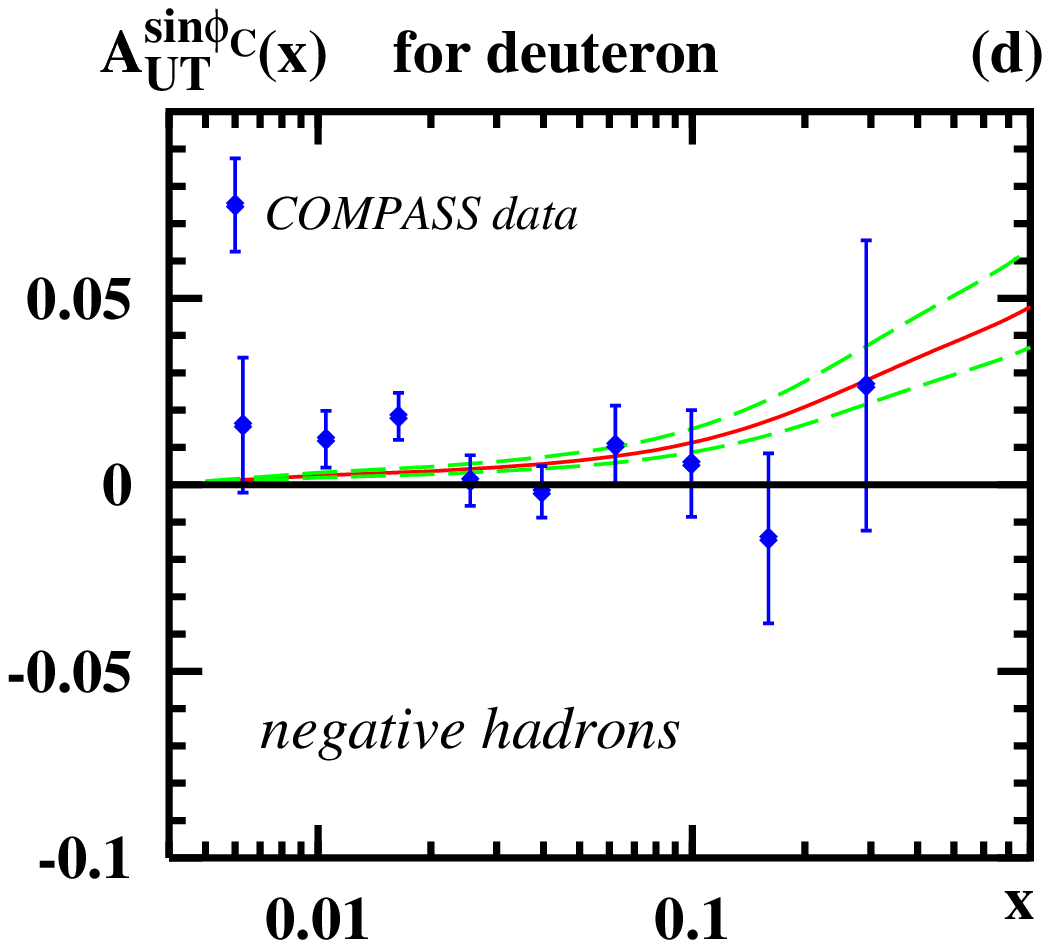}
\end{center}
\vspace{-7mm}
\caption{\label{Fig3:AUT-x}\footnotesize
Collins SSA $A_{UT}^{\sin(\phi+\phi_S)}$ as function of $x$ vs.\
HERMES \cite{Diefenthaler:2005gx} and new COMPASS
\cite{Alexakhin:2005iw} data.}
\end{figure}

The absolute numbers for $\la 2 B_{\rm G}H_1^{\rm fav}\ra$ and
$\la 2 B_{\rm G}H_1^{\rm unf}\ra$ are of similar magnitude. This
can be understood in the string fragmentation picture 
and the Sch\"afer-Teryaev sum rule \cite{Artru:1995bh}.
Fit (\ref{Eq:Apower-HERMES}) describes the HERMES proton
target data \cite{Diefenthaler:2005gx} on the Collins SSA 
(Figs.~\ref{Fig3:AUT-x}a, b) and is in agreement with COMPASS
deuteron data \cite{Alexakhin:2005iw} (Figs.~\ref{Fig3:AUT-x}c, 
d).

%

%
\begin{wrapfigure}[13]{HR}{0.35\textwidth}
\vspace{-6mm}
\begin{flushright}
\includegraphics[width=0.3\textwidth]{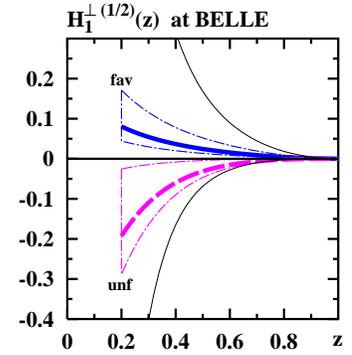}
\vspace{-2mm}\hfil
\begin{minipage}{0.3\textwidth}
\caption{\label{Fig5:BELLE-best-fit}\footnotesize
Collins FF $H_1^\aH(z)$ needed to explain
the BELLE data \cite{Abe:2005zx}.
The shown 1-$\sigma$ error bands are correlated.}
\end{minipage}
\end{flushright}
\end{wrapfigure}

\paragraph{3.2 Collins effect in $\bf e^+e^-$.}
The $\cos2\phi$ dependence of the cross section
(\ref{Eq:A1-in-e+e}) could arise also from hard gluon radiation or
detector acceptance effects. These effects, being flavor
independent, cancel out from the double ratio of $A_1^U$, where
both hadrons $h_1h_2$ are pions of unlike sign, to $A_1^L$, where
$h_1h_2$ are pions of like sign, i.e.
\be
\label{Eq:double-ratio}
\frac{A_1^U}{A_1^L}\approx 1 + \cos(2\phi_1)P_{U/L}(z_1,z_2)\;.
\ee

The BELLE data \cite{Abe:2005zx} can be described with the following
Ansatz and  best fit, which is shown in Fig. \ref{Fig5:BELLE-best-fit},
\be
\label{Eq:best-fit-BELLE}
\!\!\!H_1^\aH(z) = C_a \,z\,D_1^a(z),\;\,
C_{\rm fav}=0.15,\;\,C_{\rm unf}=-0.45.
\ee
Other Ans\"atze gave less satisfactory fits.
%
%
\begin{figure}[b!]
\begin{center}
\begin{tabular}{cccc}
\vspace{-16mm} &&& \cr
\hspace{-5mm}\includegraphics[width=1.6in]{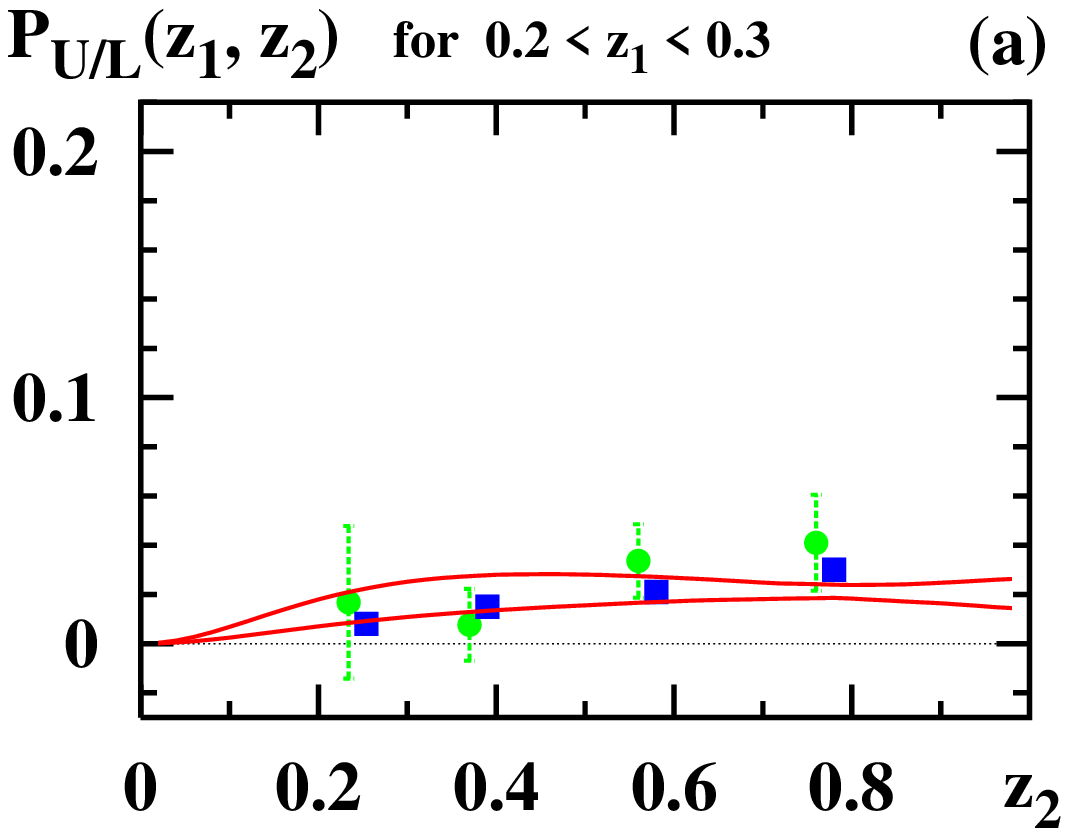} &
\hspace{-5mm}\includegraphics[width=1.6in]{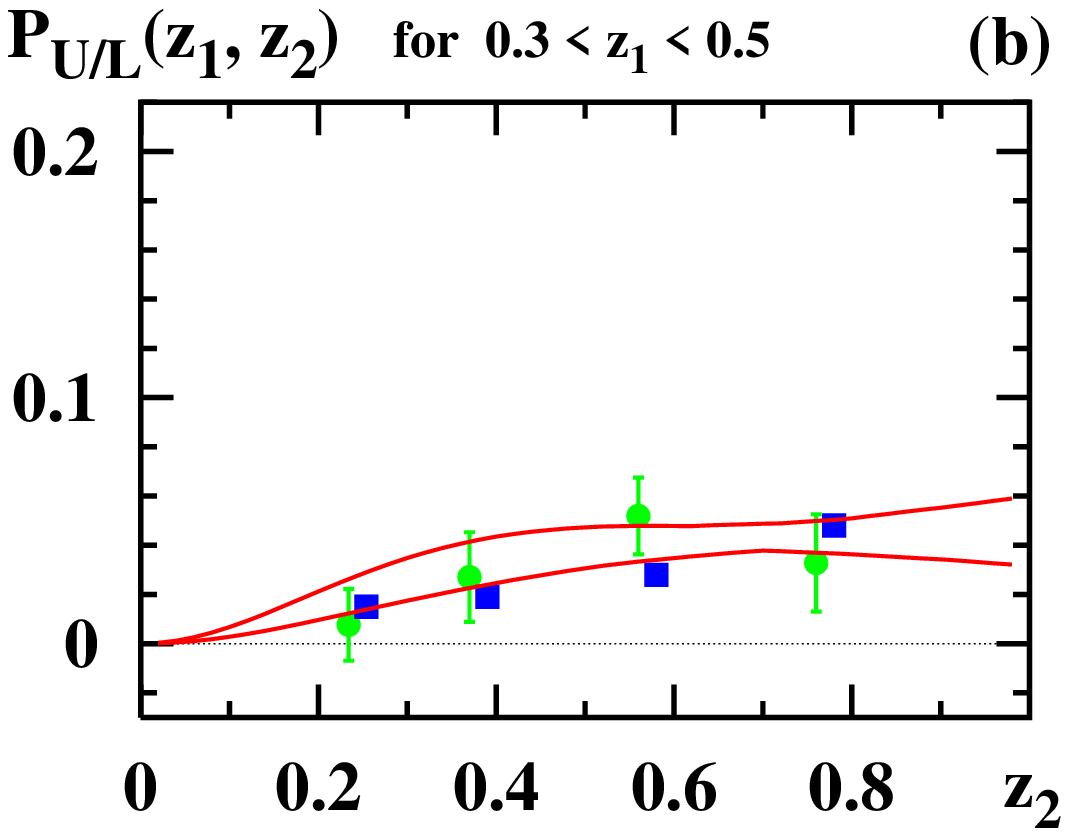} &
\hspace{-5mm}\includegraphics[width=1.6in]{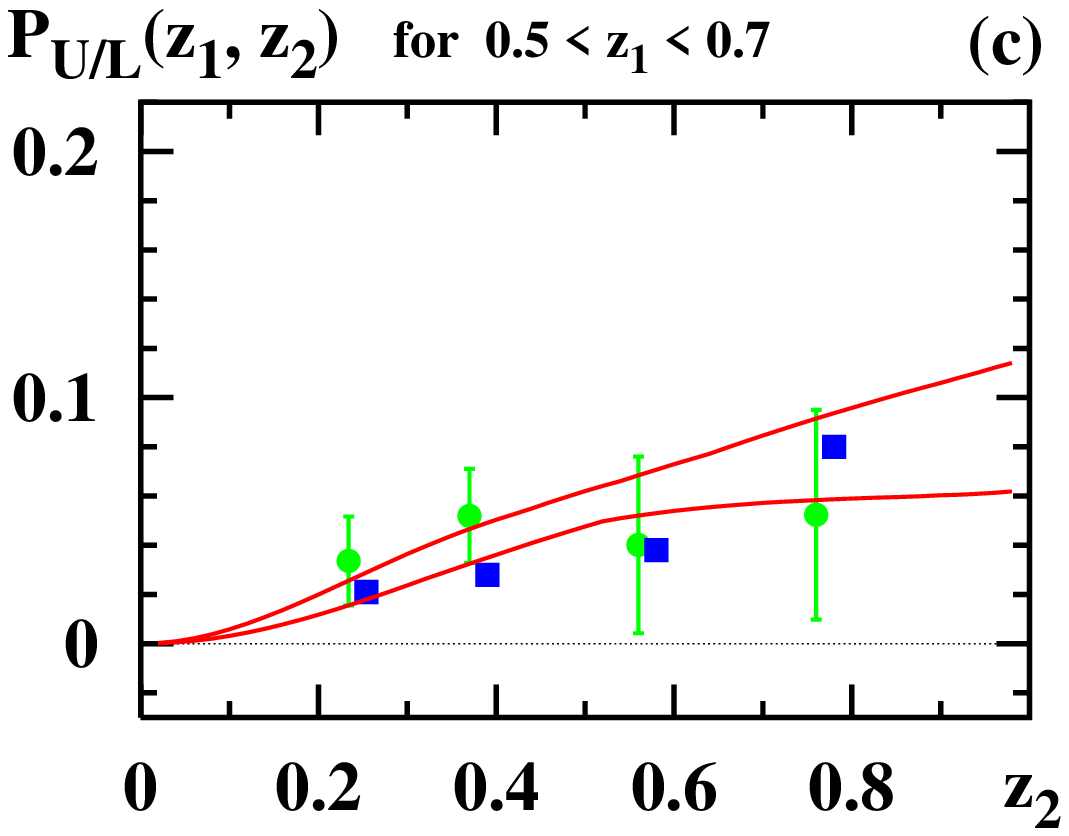} &
\hspace{-5mm}\includegraphics[width=1.6in]{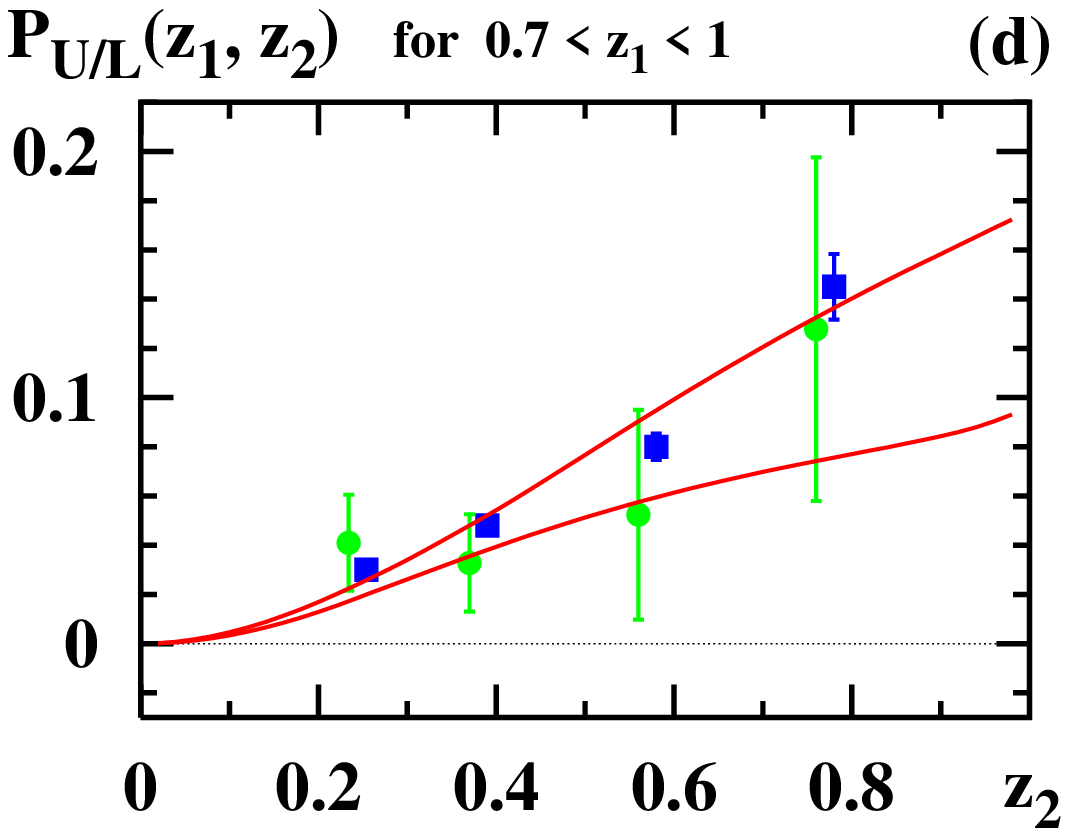} \cr
\hspace{-5mm}\includegraphics[width=1.6in]{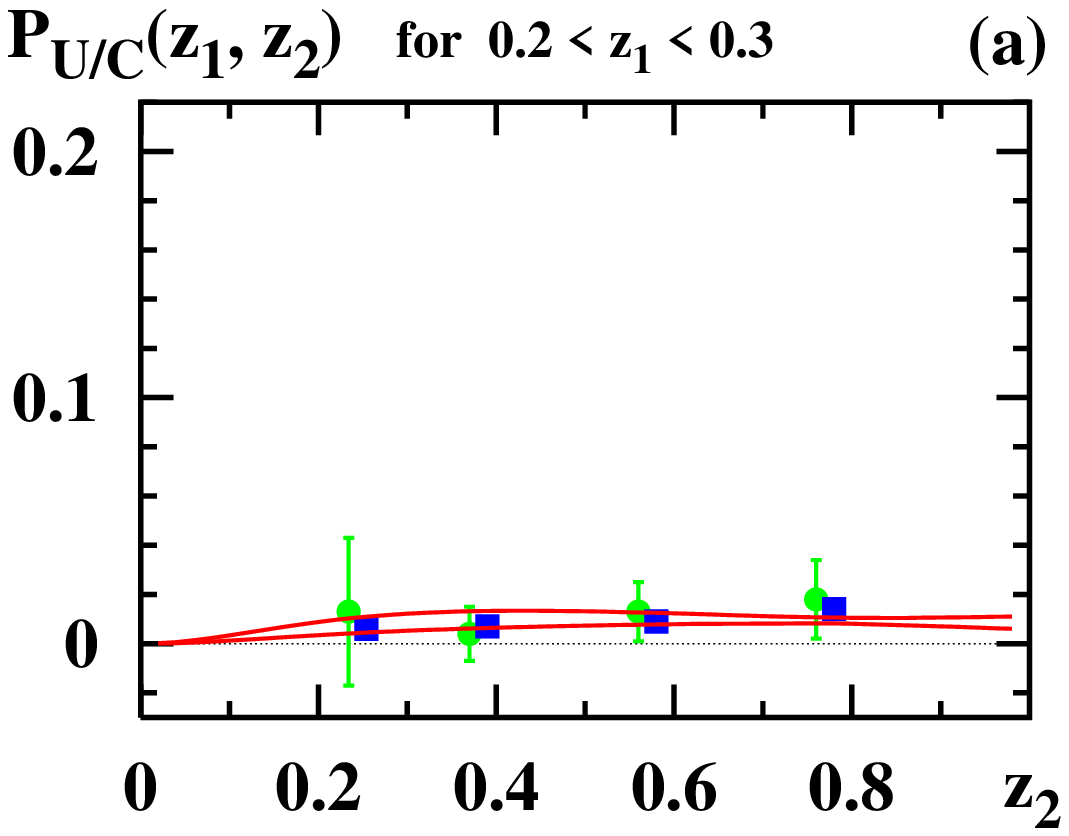} &
\hspace{-5mm}\includegraphics[width=1.6in]{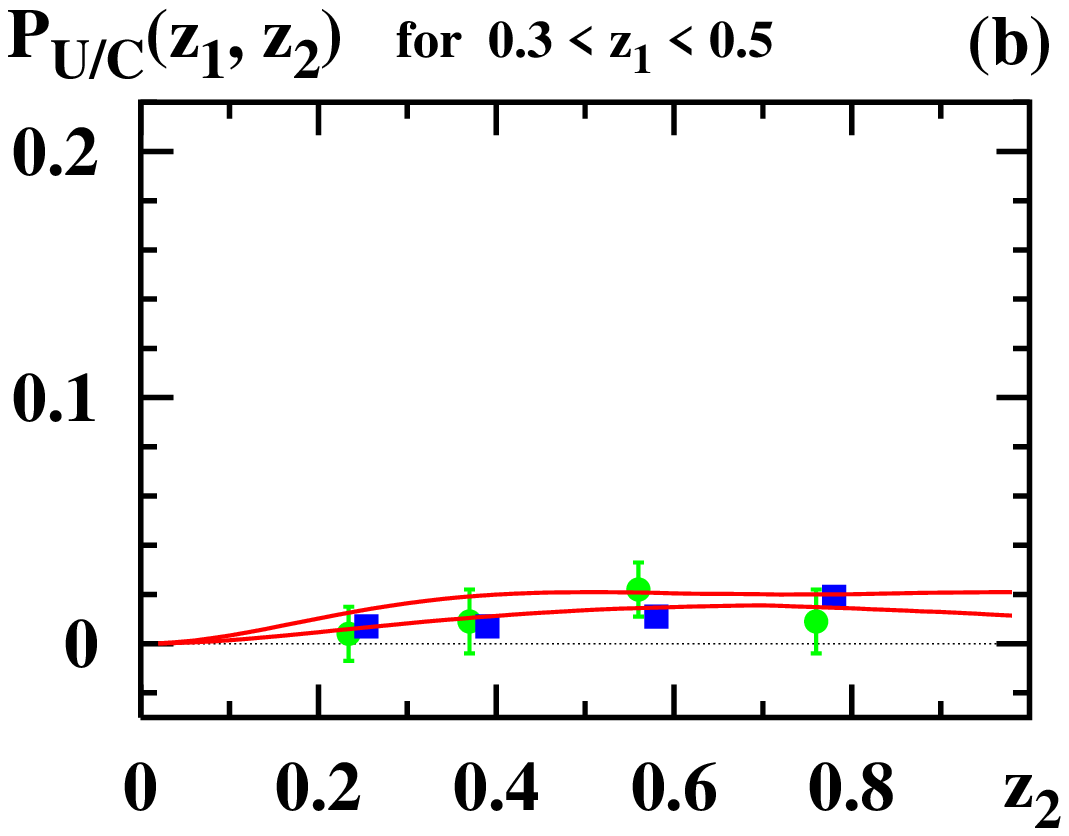} &
\hspace{-5mm}\includegraphics[width=1.6in]{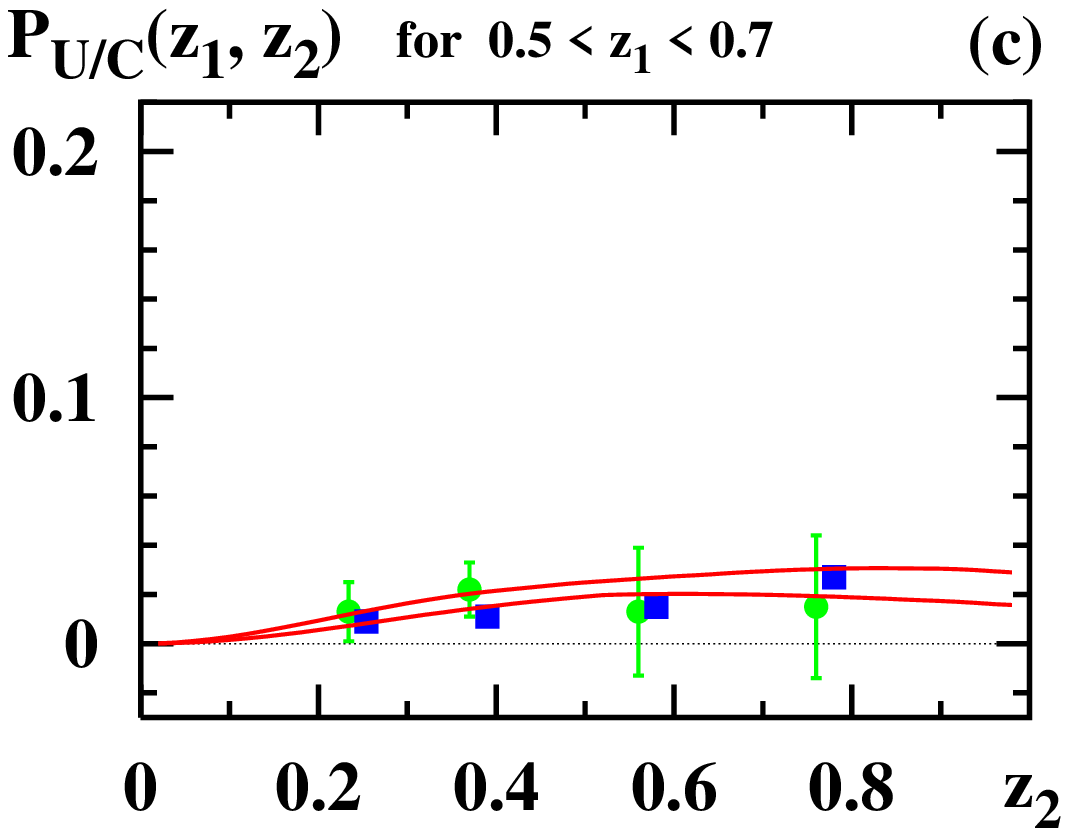} &
\hspace{-5mm}\includegraphics[width=1.6in]{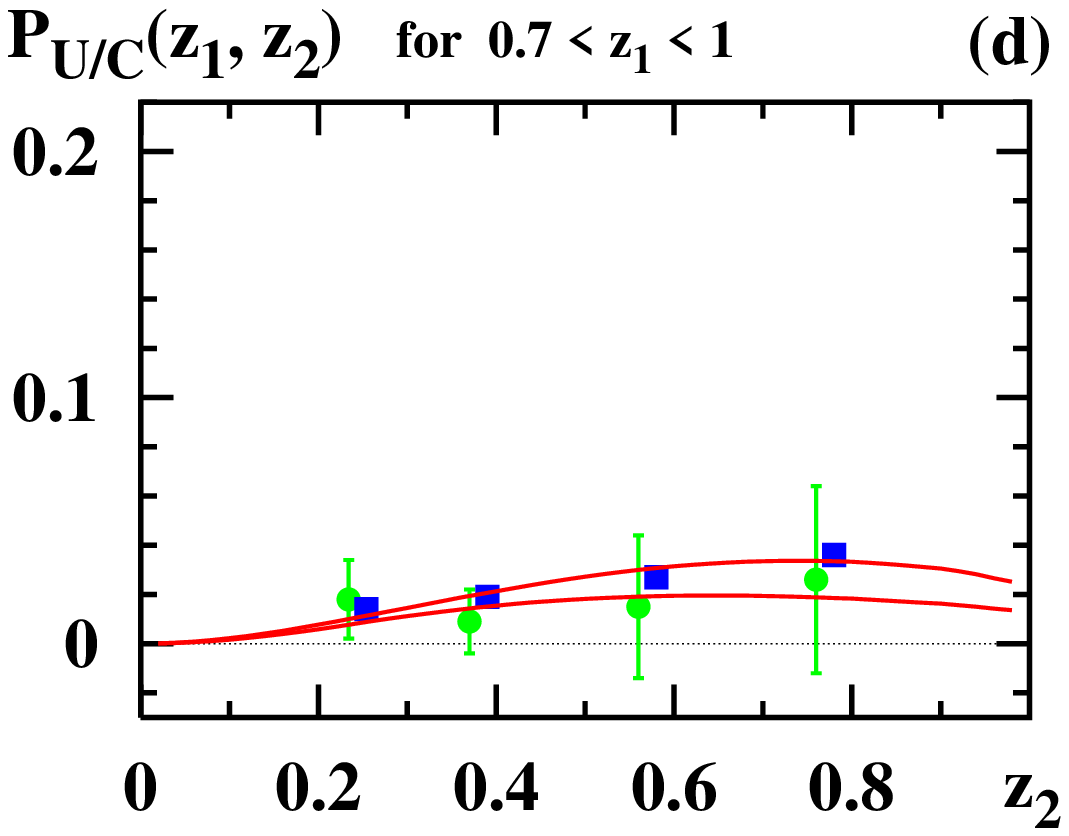}
\end{tabular}
\end{center}
\vspace{-8mm}
\caption{\label{Fig6:BELLE}\footnotesize
    {\bf Top}:
    $\;\,P_{U/L}(z_1,z_2)$ defined in
    Eq.~(\ref{Eq:double-ratio}) for fixed $z_1$-bins
    as function of $z_2$ vs.\ BELLE data~\cite{Abe:2005zx}.
    {\bf Bottom}:
    The observable $P_{U/L}(z_1,z_2)$ defined analogously, see text,
    vs. preliminary BELLE data \cite{Ogawa:2006bm}. Blue squares are 
    new preliminary data, see Sec.~\ref{Sec:new-results}.}
\end{figure}
%
The azimuthal observables in $e^+e^-$-anni\-hilation are
bilinear in $H_1^\aH$ and therefore symmetric with respect to the
exchange of the signs of $H_1^{\rm fav}$ and $H_1^{\rm unf}$.
The BELLE data \cite{Abe:2005zx} unambiguously indicate that
$H_1^{\rm fav}$ and $H_1^{\rm unf}$ have opposite signs, but they
cannot tell us which is positive and which is negative. The
definite signs in (\ref{Eq:best-fit-BELLE}) and Fig.
\ref{Fig5:BELLE-best-fit} are dictated by SIDIS data
\cite{Diefenthaler:2005gx} and model \cite{Schweitzer:2001sr}
with $h_1^u(x)>0$.
In Fig.~\ref{Fig6:BELLE} (top) the BELLE data \cite{Abe:2005zx} are
compared to the theoretical result for $P_{U/L}(z_1,z_2)$.

\newpage
\paragraph{3.3 BELLE vs.~HERMES.}
In order to compare Collins effect in SIDIS at HERMES
\cite{Airapetian:2004tw,Diefenthaler:2005gx} and in
$e^+e^-$-annihilation at BELLE \cite{Abe:2005zx} we consider the
ratios $H_1^a/D_1^a$ which might be less scale dependent. The
BELLE fit in Fig.~\ref{Fig5:BELLE-best-fit} yields in the HERMES
$z$-range:
\be\label{Eq:Apower-BELLE}
    \frac{\la 2H_1^{\rm fav}\ra}{\la D_1^{\rm fav}\ra}
    \biggr|_{\rm BELLE}\hspace{-9mm} = (5.3\cdots 20.4)\%,\quad
    \frac{\la 2H_1^{\rm unf}\ra}{\la D_1^{\rm
    unf}\ra} \biggr|_{\rm BELLE}\hspace{-9mm} =
    -(3.7\;\cdots\;41.4)\%  \;.
\ee

%
\begin{wrapfigure}[13]{R}{2.8in}
\vspace{-8mm}
\begin{flushright}
\includegraphics[width=1.25in]{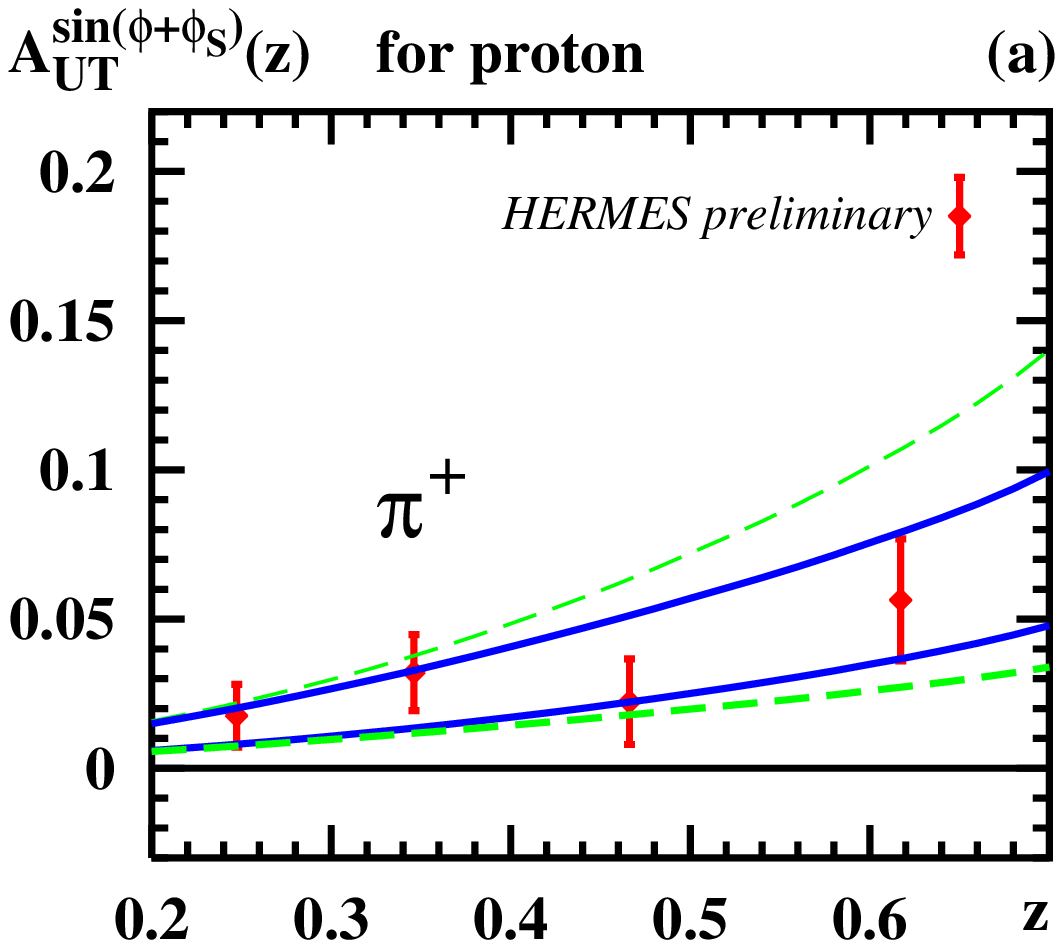}
\includegraphics[width=1.25in]{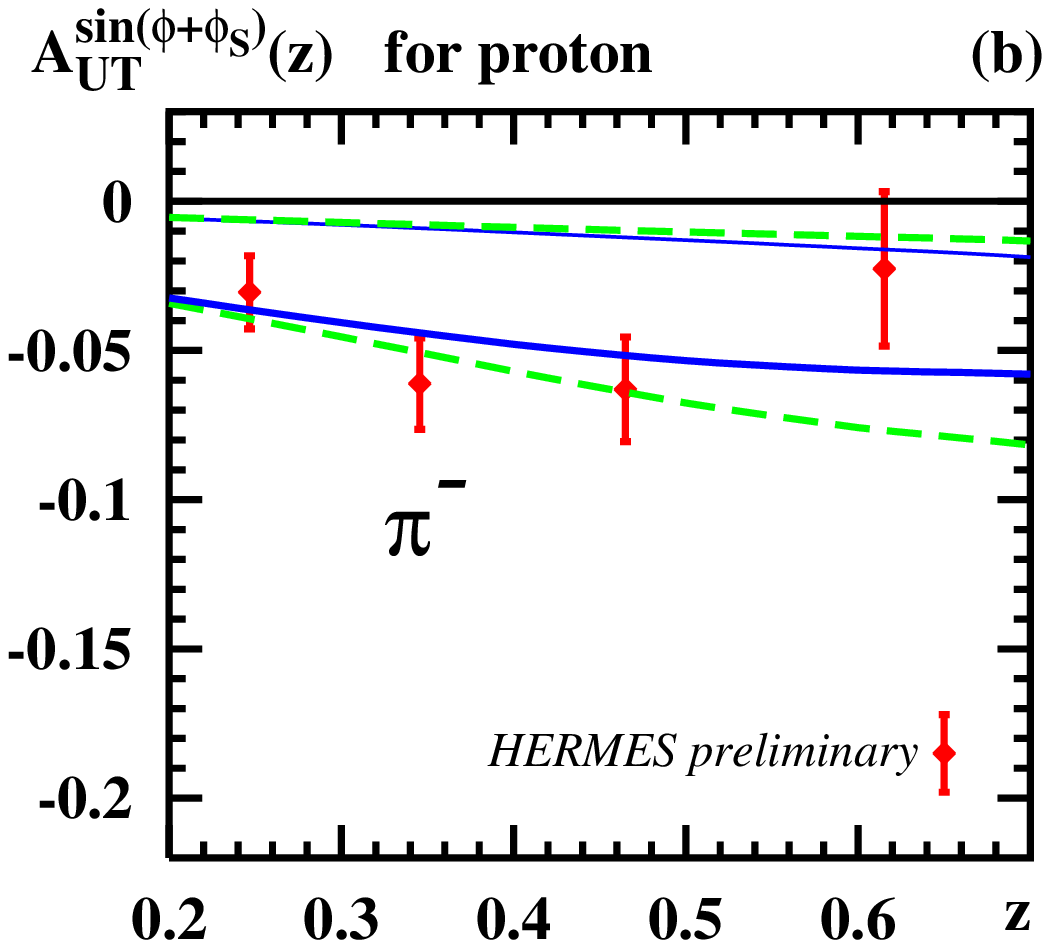}
\end{flushright}
\vspace{-10mm}
\begin{flushright}
\begin{minipage}{2.5in}
\vskip2mm
\caption{\label{Fig7:HERMES-AUT-z-from-BELLE} \footnotesize
    The Collins SSA $A_{UT}^{\sin(\phi+\phi_S)}(z)$ as function
    of $z$. The theoretical curves are based on the fit of
    $H_1^a(z)$ to the BELLE data under the assumption
    (\ref{Eq:assume-weak-scale-dep}).
    The dashed lines indicate
    the sensitivity of the SSA to the 
    Gaussian widths.
    }
\end{minipage}
\end{flushright}
\end{wrapfigure}
%

The above numbers (errors are correlated!) and the
result in Eq.~(\ref{Eq:Apower-HERMES}) are compatible,
if one takes into account the factor $B_{\rm G}<1$ in 
Eq.~(\ref{Eq:Apower-HERMES}).

Assuming a weak scale-dependence also for 
\be\label{Eq:assume-weak-scale-dep}
    \frac{H_1^\aH(z)}{D_1^a(z)}\biggr|_{\rm BELLE}
    \approx\;
    \frac{H_1^\aH(z)}{D_1^a(z)}\biggr|_{\rm HERMES}
\ee
and considering the 1-$\sigma$ uncertainty of the BELLE fit in
Fig. \ref{Fig5:BELLE-best-fit} and the sensitivity to unknown
Gaussian widths of $H_1^a(z)$ and $h_1^a(x)$, c.f.\  Footnote~1
and Ref.~\cite{Efremov:2006qm}, one obtains also a satisfactory
description of the $z$-dependence of the HERMES data
\cite{Diefenthaler:2005gx} as shown in
Fig.~\ref{Fig7:HERMES-AUT-z-from-BELLE}.

These observations allow to draw the conclusion that
it is, in fact, the same Collins effect at work in SIDIS
\cite{Airapetian:2004tw,Alexakhin:2005iw,Diefenthaler:2005gx}
and in $e^+e^-$-annihilation \cite{Abe:2005zx,Ogawa:2006bm}.
Estimates indicate that the early preliminary DELPHI result
\cite{Efremov:1998vd} is compatible with
these findings \cite{Efremov:2006qm}.

%
\begin{wrapfigure}[11]{HR}{2.1in}
\vspace{-5mm}
\centerline{\includegraphics[width=1.7in]{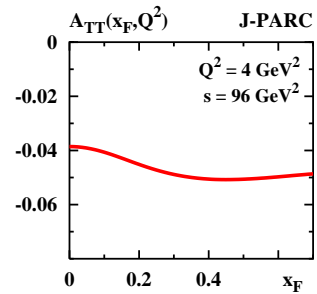}}
\vspace{-5mm}\hfil
\begin{minipage}{1.9in}
\vskip6mm
\caption{\label{Fig-DY-ATT-pp-JPARC}\footnotesize
Double spin asymmetry $A_{TT}$ in DY, Eq.~(\ref{Eq:DY}), vs.\
$x_F$ for the kinematics of J-PARC.}
\end{minipage}
\end{wrapfigure}
\paragraph{3.3 Transversity in Drell-Yan process.}
The double-spin asymmetry observable in Drell-Yan (DY) lepton-pair
production in proton-proton collisions is given in LO by
\be\label{Eq:DY}
    A_{TT}(x_F) =
    \frac{\sum_a e_a^2 h_1^a(x_1) h_1^{\bar a}(x_2)}
         {\sum_a e_a^2 f_1^a(x_1) f_1^{\bar a}(x_2)}
\ee
where $x_F=x_1-x_2$ and $x_1x_2=\frac{Q^2}{s}$. In the kinematics
of RHIC $A_{TT}$ is small and difficult to measure.

In the J-PARC experiment with $E_{\rm beam}=50\,{\rm GeV}$
$A_{TT}$ would reach $-5\,\%$ in the model
\cite{Schweitzer:2001sr}, see Fig. \ref{Fig-DY-ATT-pp-JPARC}, and
could be measured \cite{J-PARC-proposal}. The situation is
similarly promising in proposed polarized beam U70-experiment
\cite{Abramov:2005mk}.

Finally, in the PAX-experiment proposed at GSI \cite{PAX}
in polarized $\bar pp$ collisions one may expect
$A_{TT}\sim(30\cdots50)\%$ \cite{Efremov:2004qs}.
There $A_{TT}\propto h_1^u(x_1) h_1^{\bar u}(x_2)$ to a good approximation,
due to $u$-quark ($\bar u$-quark) dominance in the proton (anti-proton)
\cite{Efremov:2004qs}.

\section{New data and developements}
\label{Sec:new-results}

Since our studies were completed 
\cite{Efremov:2004tp,Collins:2005ie,Efremov:2006qm}
new data became available from SIDIS at HERMES \cite{Diefenthaler:2006vn}
and $e^+e^-$-annihilations at BELLE \cite{Ogawa:2006bm}.
What is the impact of the new experimental results?
Do they confirm our current understanding
of the Sivers- and Collins-effects, or will they require a revision?

\paragraph{4.1 New results from BELLE.}
Interesting recent news are the preliminary BELLE data
\cite{Ogawa:2006bm} for the ratio of azimuthal asymmetries of
unlike sign pion pairs, $A_1^U$, to all charged pion pairs,
$A_1^C$. The new observable $P_{U/C}$ is defined analogously to $P_{U/L}$
in Eq.~(\ref{Eq:double-ratio}) as $A_1^U/A_1^C \approx
1+\cos(2\phi)\,P_{U/C}$. 
Fig.~\ref{Fig6:BELLE} (bottom) shows that the fit (\ref{Eq:best-fit-BELLE}) 
from \cite{Efremov:2006qm} ideally describes the new experimental points!
Thus, the new data confirm the picture of the Collins function in
Fig.~\ref{Fig5:BELLE-best-fit}, but will allow to reduce the uncertainty 
of the extraction.

\paragraph{4.2 \boldmath $\pi^0$ Collins SSA.}
The (unpolarized or Collins) fragmentation functions for neutral pions
are just the average of the favoured and unfavoured fragmentation functions
into charged pions, due to isospin symmetry.
Since in the HERMES kinematics the favoured and unfavoured
Collins functions are of opposite sign and nearly equal in magnitude,
$\la 2 B_{\rm G}H_1^{\rm fav}\ra\approx-\la 2 B_{\rm G}H_1^{\rm unf}\ra$
c.f.\ Sec.~3.1, one expects the $\pi^0$ Collins SSA to be nearly zero
\cite{Efremov:2006qm}. Most recent HERMES data confirm this prediction
within error bars \cite{Diefenthaler:2006vn}.

\begin{wrapfigure}[13]{R}{3.0in}
\vspace{-5mm}
\begin{flushright}
\hspace{-35mm}\includegraphics[width=1.6in]{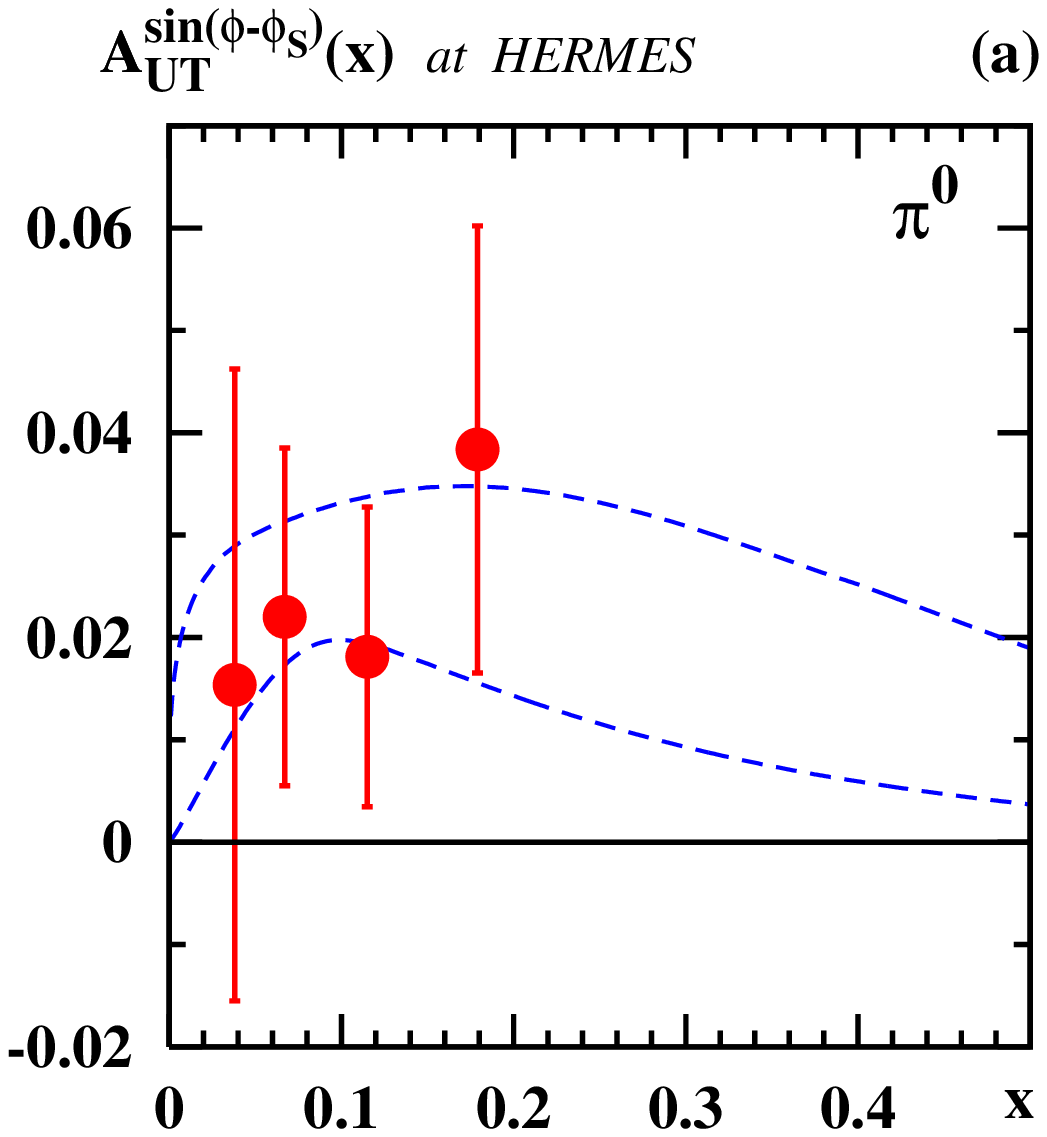}
\hspace{-5mm} \includegraphics[width=1.6in]{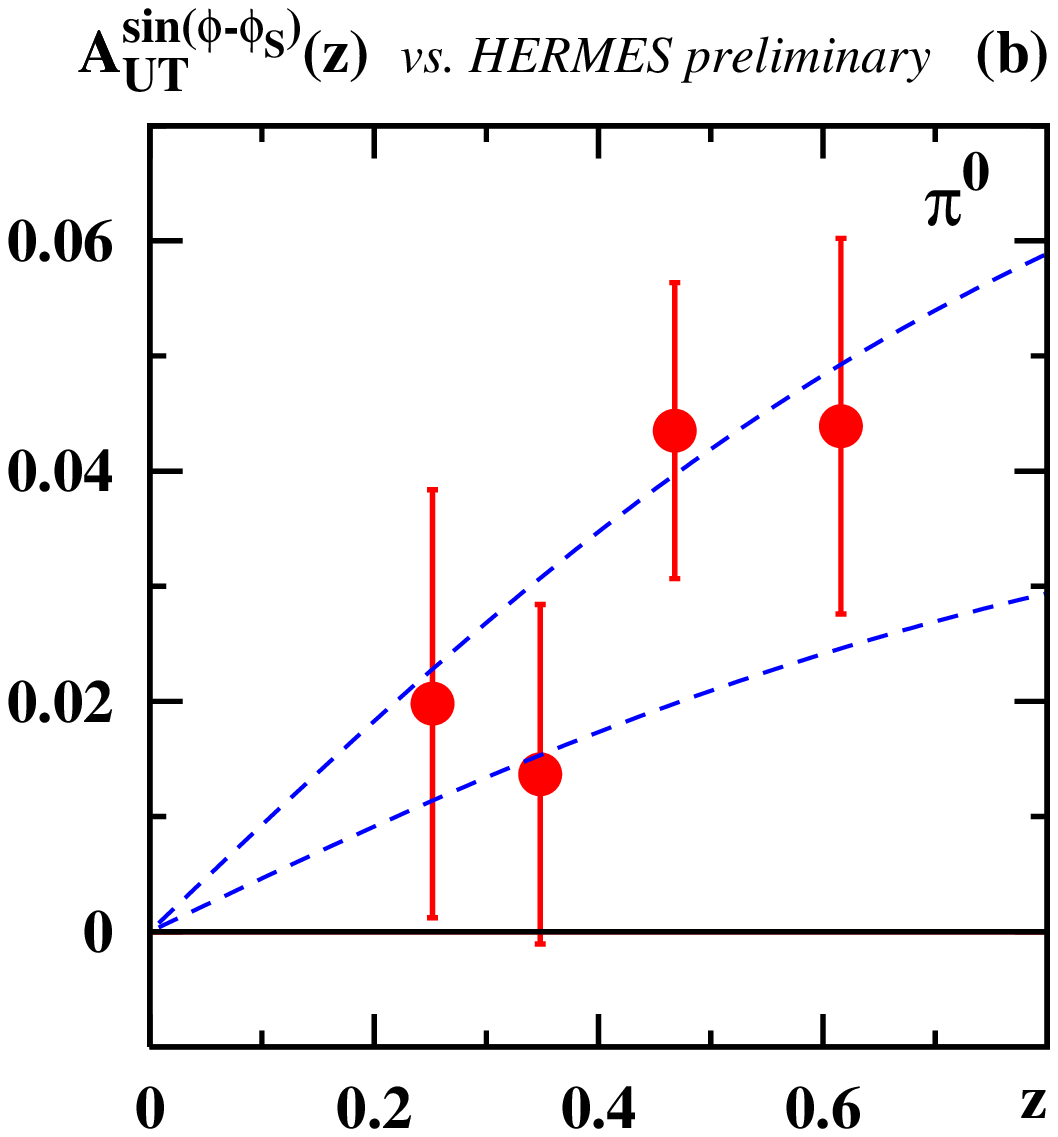}
\end{flushright}
\vspace{-12mm}
\begin{flushright}
\begin{minipage}{2.95in}
\vskip2mm
\caption{\label{Fig9:Sivers-Pi0} \footnotesize
The Sivers SSA $A_{UT}^{\sin(\phi+\phi_S)}(z)$ for $\pi^0$ as 
functions of $x$ and $z$. The preliminary HERMES data are from 
\cite{Diefenthaler:2006vn}. The theoretical curves are based on 
the extraction of the Sivers effect \cite{Collins:2005ie} from 
the HERMES data on $\pi^\pm$ SSAs \cite{Airapetian:2004tw}.}
\end{minipage}
\end{flushright}
\end{wrapfigure}
\paragraph{4.3 \boldmath $\pi^0$ Sivers SSA.}
Isospin symmetry applies not only to fragmenation functions
but to the entire effects. Thus, knowing the Sivers SSAs for
charged pions one is able to predict the effect for $\pi^0$.
In Figs.~\ref{Fig9:Sivers-Pi0}a, b we compare our predictions made on the basis
of the results from \cite{Collins:2005ie} discussed in Sec.~2.1
with the most recent HERMES data \cite{Diefenthaler:2006vn}.
The agreement is satisfactory. In particular, data on the 
$z$-dependence of the Sivers SSA provide a direct test
of the Gauss model for transverse parton and hadron momenta
\cite{Collins:2005ie}. As can be seen in Fig.~\ref{Fig9:Sivers-Pi0}b,
within the present precision of data the Gauss Ansatz is useful.

\paragraph{4.4 Sivers effect for kaons.}
In the HERMES experiment also the Sivers effect for charged kaons 
was measured. For $K^-$ the effect is compatible with zero within 
error bars. But for $K^+$ in the region of $x = (0.05-0.15)$ the 
SSA is about (2-3) times larger than the $\pi^+$ SSA 
\cite{Diefenthaler:2006vn}, while for $x\geq 0.15$ 
the $K^+$ and $\pi^+$ SSAs are of  comparable size within error bars.
Can one understand this behaviour?

The ``only difference'' between the $\pi^+$ and $K^+$ SSAs is 
the exchange $\bar d\leftrightarrow \bar s$. Therefore, in our
approach of Sec.~2.1, where we neglect the effects of Sivers strange 
and antiquarks one expects $\pi^+$ and $K^+$ SSAs of same magnitude.
However, by including explicitly $\bar u,\;\bar d,\; s$ 
and $\bar s$ Sivers distributions one could explain the observed 
enhancement of the $K^+$ Sivers SSA with respect to the $\pi^+$ SSA, 
provided the Sivers seaquark distributions would reach about $50\%$ 
of the magnitude of the Sivers quark distributions.
At small $x$ this could be a reasonable scenario,
see \cite{Efremov:2007kj} for a detailed discussion.
A simultaneous refitting of pion and kaon SSAs will give us a conclusive 
answer (see, however, the talk by Prokudin \cite{Anselmino:2007fs}).

\section{Conclusions}
Within the uncertainties of our study we find that the SIDIS data
from HERMES \cite{Airapetian:2004tw,Diefenthaler:2005gx} and
COMPASS \cite{Alexakhin:2005iw} on the Sivers and Collins SSA from
different targets are in agreement with each other and with BELLE
data on azimuthal correlations in $e^+e^-$-annihilations.

At the present stage of art large-$N_c$ predictions 
for the flavour dependence of the Sivers function 
are compatible with data, and provide useful
constraints for their analysis.

The favored and unfavored Collins FFs appear to be of comparable 
magnitude but have opposite signs, and $h_1^u(x)$ seems close to 
saturating the Soffer bound, other $h_1^a(x)$ are hardly
constrained. This conclusion is supported by a simultanuous analysis of
HERMES, COMPASS and BELLE data \cite{Anselmino:2007fs} with additional
conclusion on the tendency of $h_1^d(x)$ to be negative.
These findings are in agreement with old DELPHI
and with the most recent BELLE data
and with independent theoretical studies
\cite{Vogelsang:2005cs}. 

New HERMES and BELLE data confirm our first understanding of these
effects, except for the HERMES data on the kaon Sivers SSA which may 
provide new interesting information on Sivers seaquarks.
Further data from SIDIS (COMPASS, JLAB, HERMES) and $e^+e^-$ colliders 
(BELLE) will help to improve this first picture.

The understanding of the novel functions $f_{1T}^{\perp a},\,h_1^a$
and $H_1^a$ emerging from SIDIS and $e^+e^-$-annihilations,
however, will be completed  only thanks to future data 
spin asymmetries in the Drell-Yan process. Experiments 
are in progress or planned at RHIC, J-PARC, COMPASS, U70 and PAX 
at GSI.

While the Sivers and Collins effects are the most prominent effects,
it is important to keep in mind that there are further equally
fascinating effects to be explored 
\cite{Kotzinian:1995,Kotzinian:2006dw,Avakian:2007mv}.
Preliminary COMPASS results on compatible with zero deuteron target SSAs 
beyond the Sivers and Collins effects were presented in \cite{Kotzinian:2007uv}.


\paragraph{Acknowledgments.}
This work is  supported by BMBF (Verbundforschung), COSY-J\"ulich
project, the Transregio Bonn-Bochum-Giessen, and is part of the by
EIIIHT project under contract number RII3-CT-2004-506078. A.E. is
also supported by RFBR grant 06-02-16215, by RF MSE
RNP.2.2.2.2.6546 (MIREA) and by the Heisenberg-Landau Program of
JINR.



\footnotesize
\begin{thebibliography}{99}
\itemsep-1mm 
\bibitem{Bunce:1976yb}
  G.~Bunce {\it et al.},
  Phys.\ Rev.\ Lett.\  {\bf 36} (1976) 1113.
  
\bibitem{Apokin:1988sn}
  V.~D.~Apokin {\it et al.},
  Sov.\ J.\ Nucl.\ Phys.\  {\bf 49} (1989) 103
  [Yad.\ Fiz.\  {\bf 49} (1989) 165].

\bibitem{Adams:1991rw}
  D.~L.~Adams {\it et al.}, 
  Phys.\ Lett.\ B {\bf 261}, 201 and {\bf 264}, 462 (1991).

\bibitem{Anselmino:2002mx}
  M.~Anselmino,
  Czech.\ J.\ Phys.\  {\bf 52} (2002) C13
  [arXiv:hep-ph/0201150].

\bibitem{Airapetian:2004tw}
  A.~Airapetian {\it et al.}  [HERMES Collaboration],
  Phys.\ Rev.\ Lett.\  {\bf 94} (2005) 012002.

\bibitem{Alexakhin:2005iw}
  V.~Y.~Alexakhin {\it et al.}  [COMPASS Collaboration],
  Phys.\ Rev.\ Lett.\  {\bf 94} (2005) 202002.\\
  E.~S.~Ageev {\it et al.}  [COMPASS Collaboration],
  Nucl.\ Phys.\  B {\bf 765} (2007) 31.

\bibitem{Diefenthaler:2005gx}
  M.~Diefenthaler,
  AIP Conf.\ Proc.\  {\bf 792} (2005) 933
  [arXiv:hep-ex/0507013].

\bibitem{Abe:2005zx}
  K.~Abe {\it et al.}  [BELLE Collaboration],
  {Phys.\ Rev.\ Lett.} {\bf 96} { (2006)} { 232002}.

\bibitem{Ji:2004wu}
  X.~D.~Ji, J.~P.~Ma and F.~Yuan,
  Phys.\ Rev.\ D {\bf 71} (2005) 034005;
  Phys.\ Lett.\ B {\bf 597} (2004) 299.
  J.~C.~Collins and A.~Metz,
  Phys.\ Rev.\ Lett.\  {\bf 93} (2004) 252001.

\bibitem{Mulders:1995dh}
  P.~J.~Mulders and R.~D.~Tangerman,
  Nucl.\ Phys.\ B {\bf 461} (1996) 197.\\
  D.~Boer and P.~J.~Mulders,
  Phys.\ Rev.\ D {\bf 57} (1998) 5780
  [arXiv:hep-ph/9711485].

\bibitem{Boer:1997mf}
  D.~Boer, R.~Jakob and P.~J.~Mulders,
  {Nucl. Phys.} {B\bf 504} { (1997)} { 345};
  {Phys.\ Lett.} {B\bf 424} { (1998)} { 143}.


\bibitem{Sivers:1989cc}
  D.~W.~Sivers,
  Phys.\ Rev.\ D {\bf 41} (1990) 83;
  Phys.\ Rev.\ D {\bf 43} (1991) 261.

\bibitem{Brodsky:2002cx}
  S.~J.~Brodsky, D.~S.~Hwang, I.~Schmidt,
  Phys.\ Lett.\ B {\bf 530} (2002) 99;
  Nucl.\ Phys.\ B {\bf 642} (2002) 344.

\bibitem{Collins:2002kn}
  J.~C.~Collins,
  Phys.\ Lett.\ B {\bf 536} (2002) 43. 

\bibitem{Belitsky:2002sm}
  A.~V.~Belitsky, X.~Ji and F.~Yuan,
  Nucl.\ Phys.\ B {\bf 656} (2003) 165.
  X.~D.~Ji and F.~Yuan,
  Phys.\ Lett.\ B {\bf 543} (2002) 66.
  D.~Boer, P.~J.~Mulders and F.~Pijlman,
  Nucl.\ Phys.\ B {\bf 667} (2003) 201.

 \bibitem{Collins:1992kk}
  J.~C.~Collins,
  Nucl.\ Phys.\ B {\bf 396} (1993) 161 [arXiv:hep-ph/9208213].\\
  A.~V.~Efremov, L.~Mankiewicz and N.~A.~Tornqvist,
  Phys.\ Lett.\ B {\bf 284} (1992) 394.

\bibitem{Anselmino:2004ky}
  M.~Anselmino, M.~Boglione, U.~D'Alesio, E.~Leader and F.~Murgia,
  Phys.\ Rev.\ D {\bf 71} (2005) 014002.
  B.~Q.~Ma, I.~Schmidt and J.~J.~Yang,
  Eur.\ Phys.\ J.\ C {\bf 40} (2005) 63.

\bibitem{Efremov:2004tp}
  A.~V.~Efremov, K.~Goeke, S.~Menzel, A.~Metz and P.~Schweitzer,
  Phys.\ Lett.\ B {\bf 612} (2005) 233.

\bibitem{Anselmino:2005ea}
  M.~Anselmino, M.~Boglione, U.~D'Alesio, A.~Kotzinian, F.~Murgia
  and A.~Prokudin,
  Phys.\ Rev.\ D {\bf 72} (2005) 094007.

\bibitem{Vogelsang:2005cs}
  W.~Vogelsang and F.~Yuan,
  Phys.\ Rev.\ D {\bf 72} (2005) 054028
  [arXiv:hep-ph/0507266].

\bibitem{Collins:2005ie}
  J.~C.~Collins, A.~V.~Efremov, K.~Goeke, S.~Menzel, A.~Metz and P.~Schweitzer,
  Phys.\ Rev.\ D {\bf 73} (2006) 014021
  [arXiv:hep-ph/0509076].
  J.~C.~Collins {\it et al.},
  arXiv:hep-ph/0510342.

\bibitem{Anselmino:2005an}
  M.~Anselmino {\it et al.},
  arXiv:hep-ph/0511017.
  
\bibitem{Efremov:2006qm}
  A.~V.~Efremov, K.~Goeke and P.~Schweitzer,
  {Phys.\ Rev.} {D\bf 73} { (2006)} { 094025}.

\bibitem{Anselmino:2007fs}
  M.~Anselmino, M.~Boglione, U.~D'Alesio, A.~Kotzinian, F.~Murgia, A.~Prokudin 
  and C.~Turk,
  Phys.\ Rev.\  D {\bf 75} (2007) 054032
  [arXiv:hep-ph/0701006]. A. Prokudin, these Proceedings.

\bibitem{Collins:2005rq}
  J.~C.~Collins {\it et al.},
  Phys.\ Rev.\ D {\bf 73} (2006) 094023;
  Czech.\ J.\ Phys.\  {\bf 56} (2006) C125.

\bibitem{Efremov:2007kj}
  A.~V.~Efremov, K.~Goeke and P.~Schweitzer,
  Czech.\ J.\ Phys.\  {\bf 56} (2006) F181
  [arXiv:hep-ph/0702155].

\bibitem{Airapetian:1999tv}
  A.~Airapetian {\it et al.}  [HERMES Collaboration],
  Phys.\ Rev.\ Lett.\  {\bf 84} (2000) 4047;
  Phys.\ Rev.\ D {\bf 64} (2001) 097101;
  Phys.\ Lett.\ B {\bf 562} (2003) 182,
  {\bf 622} (2005) 14, and
  {\bf 648} (2007) 164.

\bibitem{Avakian:2003pk}
  H.~Avakian {\it et al.}  [CLAS Collaboration],
  Phys.\ Rev.\ D {\bf 69}, 112004 (2004)
  [arXiv:hep-ex/0301005].

\bibitem{Efremov:2001cz}
  A.~V.~Efremov, K.~Goeke and P.~Schweitzer,
  Phys.\ Lett.\ B {\bf 522} (2001) 37, 
  {\bf 544} (2002) 389E.

\bibitem{Efremov:2001ia}
  A.~V.~Efremov, K.~Goeke and P.~Schweitzer,
  Eur.\ Phys.\ J.\ C {\bf 24} (2002) 407; 
  Phys.\ Lett.\ B {\bf 568} (2003) 63;
  Eur.\ Phys.\ J.\  C {\bf 32} (2003) 337.

\bibitem{Adams:2003fx}
  J.~Adams {\it et al.}  [STAR Collaboration],
  Phys.\ Rev.\ Lett.\  {\bf 92} (2004) 171801
  [arXiv:hep-ex/0310058]. G. Bunce, these Proceedings.

\bibitem{Pobylitsa:2003ty}
   P.~V.~Pobylitsa,
   arXiv:hep-ph/0301236.

\bibitem{Burkardt:2002ks}
   M.~Burkardt,
   Phys.\ Rev.\ D {\bf 66} (2002) 114005;
  Phys.\ Rev.\ D {\bf 69} (2004) 057501;
  Phys.\ Rev.\ D {\bf 69} (2004) 091501.

\bibitem{Efremov:eb}
  A.~V.~Efremov and O.~V.~Teryaev,
  Yad.\ Fiz.\  {\bf 39} (1984) 1517;
  Phys.\ Lett.\ B {\bf 150} (1985) 383.\\
  J.~W.~Qiu and G.~Sterman,
  Phys.\ Rev.\ Lett.\  {\bf 67} (1991) 2264;
  Nucl.\ Phys.\ B {\bf 378} (1992) 52.\\
  Y.~Kanazawa and Y.~Koike,
  Phys.\ Lett.\ B {\bf 478} (2000) 121;
  Phys.\ Lett.\ B {\bf 490} (2000) 99.

\bibitem{Bomhof:2004aw}
  C.~J.~Bomhof, P.~J.~Mulders and F.~Pijlman,
  Phys.\ Lett. \ B {\bf 596} (2004) 277.\\
  A.~Bacchetta, C.~J.~Bomhof, P.~J.~Mulders and F.~Pijlman,
  Phys.\ Rev.\ D {\bf 72} (2005) 034030.\\
  O.~Teryaev, these Proceedings and references therein.

\bibitem{Bravar:1999rq}
  A.~Bravar,  
  {Nucl.\ Phys.} {B\bf 79} { (1999)} { 520c}.

\bibitem{Efremov:1998vd}
  A.~V.~Efremov, O.~G.~Smirnova and L.~G.~Tkachev,
  {Nucl.\ Phys.} {B\bf 74} { (1999)} { 49c}
  [arXiv:hep-ph/9812522];
  {\it ibid.} {B\bf 79} (1999) 554.

\bibitem{Schweitzer:2001sr}
  P.~Schweitzer {\it et al.},
  {Phys. Rev.} {D\bf 64} { (2001)} { 034013}
  [arXiv:hep-ph/0101300].

\bibitem{Artru:1995bh}
  X.~Artru, J.~Czy\.zewski and H.~Yabuki,
  {Z.\ Phys.} {C\bf 73} { (1997)} { 527};
  {Acta Phys.\ Polon.} {B\bf 29} { (1998)} { 2115}
  [arXiv:hep-ph/9805463].
  A.~Sch\"afer and O.~V.~Teryaev,
  {Phys.\ Rev.} {D\bf 61} { (2000)} { 077903}.

\bibitem{J-PARC-proposal}
  D.~Dutta {\it et al.}, J-PARC Letter of Intent (2002).

\bibitem{Abramov:2005mk}
  V.~V.~Abramov {\it et al.},
  arXiv:hep-ex/0511046. A. Vasiliev, this proceedings and 
  references therein.

\bibitem{PAX}
  P.~Lenisa and F.~Rathmann  {\it et al.}  [PAX Collaboration],
  arXiv:hep-ex/0505054.\\
  F.~Rathmann {\it et al.},
  Phys.\ Rev.\ Lett.\ {\bf 94} (2005) 014801
  [arXiv:physics/0410067].

\bibitem{Efremov:2004qs}
  A.~V.~Efremov, K.~Goeke and P.~Schweitzer,
  Eur.\ Phys.\ J.\ C {\bf 35} (2004) 207;
  arXiv:hep-ph/0412427.
  M.~Anselmino, V.~Barone, A.~Drago and N.~N.~Nikolaev,
  Phys.\ Lett.\ B {\bf 594} (2004) 97.\\
  B.~Pasquini, M.~Pincetti and S.~Boffi,
  Phys.\ Rev.\  D {\bf 76} (2007) 034020
  [arXiv:hep-ph/0612094].

\bibitem{Diefenthaler:2006vn}
  M.~Diefenthaler,
  arXiv:hep-ex/0612010;
  arXiv:0706.2242 [hep-ex].
  V. Korotkov, these Proceedings.

\bibitem{Ogawa:2006bm}
  A.~Ogawa, M.~Grosse-Perdekamp, R.~Seidl and K.~Hasuko,
  arXiv:hep-ex/0607014. \\
  M.~Grosse-Perdekamp, these Proceedings and 
  references therein.

\bibitem{Kotzinian:1995}
  A.~Kotzinian, Nucl. Phys. {\bf B441}, 234 (1995).

\bibitem{Kotzinian:2006dw}
  A.~Kotzinian, B.~Parsamyan and A.~Prokudin,
  Phys.\ Rev.\  D {\bf 73} (2006) 114017.

\bibitem{Avakian:2007mv}
  H.~Avakian, A.~V.~Efremov, K.~Goeke, A.~Metz, P.~Schweitzer and T.~Teckentrup,
  arXiv:0709.3253. 

\bibitem{Kotzinian:2007uv}
  A.~Kotzinian  [on behalf of the COMPASS collaboration],
  arXiv:0705.2402 [hep-ex].

\end{thebibliography}
\end{document}